\documentclass[aps,prd,superscriptaddress,nofootinbib,11pt]{revtex4}
\usepackage[english]{babel}
\usepackage[utf8]{inputenc}
\usepackage{graphicx}   
\usepackage{slashed}
\usepackage{epstopdf}
\usepackage{verbatim}   
\usepackage{color}      
\usepackage{subfigure}  
\usepackage{multirow}
\usepackage{hyperref}   
\usepackage{float}
\usepackage{epsfig,rotating}
\usepackage{amsmath,amssymb}
\usepackage{dsfont}
\restylefloat{table}
\numberwithin{equation}{section}

\newcommand{\vx}{\vec{x}}

\newcommand{\vk}{\vec{k}}

\newcommand{\be}{\begin{equation}}
\newcommand{\ee}{\end{equation}}
\newcommand{\bea}{\begin{eqnarray}}
\newcommand{\eea}{\end{eqnarray}}

\begin{document}
\title{Gravitational production of nearly thermal    fermionic Dark Matter .}

\author{Nathan Herring}
\email{nmh48@pitt.edu} \affiliation{Department of Physics and
Astronomy, University of Pittsburgh, Pittsburgh, PA 15260}
\author{Daniel Boyanovsky}
\email{boyan@pitt.edu} \affiliation{Department of Physics and
Astronomy, University of Pittsburgh, Pittsburgh, PA 15260}

 \date{\today}

\begin{abstract}
We consider  the cosmological production of   fermionic dark matter    during inflation and a post-inflationary radiation dominated era. This fermion \emph{only interacts gravitationally},   has a mass $m$  much smaller than the Hubble scale during inflation (but is otherwise arbitrary) and is in its Bunch-Davies vacuum state during inflation. We focus on  superhorizon modes at the end of inflation, and assume instantaneous reheating. We  obtain the full energy momentum tensor discussing its renormalization, and show that the contribution from particle production is of  the kinetic-fluid form  near matter-radiation equality.  We find  {exactly} the distribution function of produced particles $|B(k)|^2=\frac{1}{2}\Big[1-(1-e^{-\frac{k^2}{2mT_H}})^{1/2}\Big]$ which exhibits an `` emergent  temperature'' $T_H=H_0\sqrt{\Omega_R}\simeq 10^{-36}(\mathrm{eV})$. The energy density of  produced particles $\rho_{pp}$ is very similar to that of a non-relativistic degree of freedom thermalized at temperature $T_H$,  $\rho_{pp}  \propto m \, (m\,T_H)^{3/2}/a^3 $ with abundance $\Omega_{pp} \simeq \big(m/10^8\,\mathrm{GeV}\big)^{5/2}$ and ``cold'' equation of state $w(a) \simeq (T_H/m a^2)$, both dominated by superhorizon modes at the end of inflation. We discuss subtle aspects of isocurvature perturbations.

\end{abstract}


\maketitle

\section{Introduction}

While the cosmological and astrophysical evidence for, and necessity of, Dark Matter (DM) is compelling, it is abundantly clear that a particle physics  candidate must be sought in extensions beyond the Standard Model.  A multi decade effort for direct detection of various possible candidates has not yet led to the identification of a (DM) particle\cite{bertone}-\cite{nowimp2}. A  theoretical challenge in proposing a suitable particle physics candidate is to identify a production mechanism that  yields the correct abundance and equation of state to satisfy the cosmological and astrophysical constraints and whose lifetime is of the order of, or larger than, the age of the Universe.

 Particle production  as a  consequence of cosmological expansion is a remarkable phenomenon that was studied in pioneering work   in refs.\cite{parker,ford,moste1,birrell,fullbook,parkerbook,mukhabook}. An important aspect of this production mechanism is that it is naturally a consequence of the dynamical gravitational background, and if the  particle only interacts with gravity and no other degrees of freedom, its abundance is determined solely by the particle mass, its coupling to gravity, and cosmological parameters, independent of hypothetical couplings beyond the Standard Model.

 Gravitational production has been studied for various candidates and different cosmological backgrounds: heavy particles produced during inflation\cite{heavydm1,heavydm2,heavydm3,kuzmin}, via inflaton oscillations\cite{vela,ema1,ema2}, reheating\cite{hash,vilja1}, or via cosmological expansion during an era with a particular equation of state\cite{vilja2}, and more recently ultralight bosonic particles cosmologically produced during inflation and a post-inflation radiation era\cite{herring}.

 The study of cosmological production of a fermionic species   has received far less attention. Early work\cite{mostafer,audretsch} addressed this important cosmological production channel within the context of standard cosmology, which was  later extended to  various inflationary scenarios\cite{lyth,chungfer,feraxion,ema,kuzmin,kuzmin2}.

 In this article we focus on studying in detail the cosmological production of a fermionic species that only interacts with gravity, setting up initial conditions during a   de Sitter inflationary era   and matching onto a post-inflation radiation dominated (RD) era, with important differences from previous studies \cite{chungfer}:

 \textbf{i:)}We consider the non-adiabatic gravitational production of a fermionic degree of freedom  throughout  the inflationary and post-inflationary radiation dominated era until matter-radiation equality. The fermion mass $m$ is taken to be much smaller than the Hubble scale during inflation, but is otherwise arbitrary.    We solve  {exactly} the Dirac equation during inflation and radiation domination with the proper boundary conditions, and match the solutions at the transition from inflation to radiation domination (RD).

  \textbf{ii:)} This fermionic degree of freedom does \emph{not} couple to the inflaton or any other field, it only interacts gravitationally,  and is in its Bunch-Davies vacuum state during inflation, which is taken to be described by a de Sitter space-time.

  \textbf{iii:)} We focus on super-Hubble wavelengths at the end of inflation, since these are the cosmologically relevant scales. Since these modes are outside the particle horizon and describe slow dynamics causally disconnected from sub-horizon microphysics,  we  assume  a rapid transition from de Sitter inflation on to a radiation dominated stage.   We obtain  {exactly} the distribution function of the produced particles, and establish consistently that the superhorizon modes at the end of inflation yield the largest contribution to the final abundance and equation of state.

   \textbf{iv:)} We \emph{do not} invoke the adiabatic approximation to obtain a particle number. Instead, we obtain the full energy momentum tensor, and its expectation value in the ``in'' Bunch-Davies vacuum state. We discuss in detail its renormalization and unambiguously  extract the contribution from particle production near matter radiation equality. We show that the asymptotic regime becomes adiabatic well before matter radiation equality and show   that in this adiabatic regime the  renormalized energy momentum tensor features the kinetic-fluid form after subtraction of the zero point contribution.

\vspace{1mm}

\textbf{Summary of main results:}

We consider one   fermionic species   in a cosmological background from de Sitter inflation followed by a radiation dominated (RD) era. This fermion has a mass $m$ much smaller than the Hubble scale during inflation, $H_{dS}$, and does not couple to any other field. We focus on wavelengths that are much larger than the particle horizon at the end of inflation; these modes describe slow evolution and are causally decoupled from the microphysics thereby justifying the assumption of  a rapid transition from the  de Sitter inflationary stage to a radiation dominated (RD) era. The fermionic field is in its Bunch-Davies vacuum state during inflation. The mode functions for the spinor solutions of the Dirac equation are found  {exactly} during inflation and (RD) with proper asymptotic boundary conditions,  and matched continuously across the transition. We consider space-time as a \emph{background}: during inflation cosmological dynamics is dominated by the inflaton, and during (RD)   by the $\gtrsim  100$ degrees of freedom of the standard model (and beyond).  Thus the (DM) contribution is negligible during these eras until near matter radiation equality.  We \emph{do not} introduce an interpolating number operator based on some adiabatic approximation; instead we obtain the exact energy momentum tensor valid during and post-inflation. We discuss in detail its renormalization and extract the contribution from particle production near matter radiation equality, when the adiabatic approximation is valid. In this regime we find that the particle production contribution to the energy momentum tensor   is of the kinetic fluid form with a distribution function of the produced particles $|B_k|^2 = \frac{1}{2}\,\Big[1-(1-e^{-\frac{k^2}{2mT_H}})^{1/2}\Big]$ with $k$ the comoving wavevector and  an \emph{emergent} temperature $T_H = \frac{H_0}{2\pi}\sqrt{\Omega_R} \simeq 10^{-36} \,\mathrm{eV}$, with $H_0,\Omega_R$ the Hubble expansion rate and radiation fraction today. This distribution function is remarkably similar to a Maxwell-Boltzmann distribution for a non-relativistic particle in thermal equilibrium at temperature $T_H$ and vanishing chemical potential.  We confirm, self-consistently, that the contribution to the abundance and equation of state is completely dominated by wavevectors that were well outside the horizon at the end of inflation. We discuss subtle aspects of isocurvature perturbations. A comparison between fermionic and bosonic fields conformally coupled to gravity is also discussed.

\vspace{1mm}

This article is organized as follows: in section (\ref{sec:model}) we introduce the model and main assumptions. In section (\ref{sec:solutions}) we  obtain the \emph{exact} fermionic spinors with ``in'' and ``out'' boundary conditions during inflation and (RD) respectively, matching them at the end of inflation, and obtain the Bogoliubov coefficients. In section (\ref{sec:tmunu}) we obtain the full energy momentum tensor, discuss in detail its renormalization and show explicitly that in the adiabatic regime the contribution from particle production features the kinetic-fluid form, with a distribution function determined by the Bogoliubov coefficients. This distribution function exhibits an ``emergent'' temperature and is remarkably similar to the Maxwell-Boltzmann distribution function for a non-relativistic degree of freedom thermalized at this   temperature. In this section we determine the (DM) abundance from this fermionic species and its equation of state parameter $w$. In section (\ref{sec:iso}) we discuss important and subtle issues associated with isocurvature perturbations in the case under study. Section (\ref{sec:discussion}) presents a discussion of various aspects and comparison with previous work and the bosonic case. Section (\ref{sec:conclusions}) summarizes our results and conclusions. Several appendices provide  various technical details.

\section{The model}\label{sec:model}
We consider a free Dirac fermion of mass $m$ as  a dark matter candidate, the generalization to Majorana fermions is   described in appendix (\ref{app:majorana}). Our main assumptions are the following:

\textbf{i:)} It  does not interact with any other field, including the inflaton or any other field that drives inflation. It only interacts gravitationally. It is light as compared to the Hubble scale during inflation $H_{dS}$, and we focus on superhorizon wavelengths at the end of inflation, since these are the most relevant for structure formation. The small dimensionless parameters $\varepsilon = \sqrt{m/H_{dS}} \ll 1$ and $k\eta_R \ll 1$ with $\eta_R$ the horizon scale at the end of inflation, furnish two small parameters that allow for an exact solution of Bogoliubov coefficients (see below).

\textbf{ii:)} The inflationary stage is described by an exact de Sitter space-time, thereby neglecting slow roll corrections, and    the fermion field is in its Bunch-Davies vacuum state during this stage.

\textbf{iii:)} We assume instantaneous reheating: namely we consider an instantaneous transition from the inflationary to a radiation dominated stage post-inflation. There is as yet an incomplete understanding of the non-equilibrium dynamics of reheating. Reheating dynamics  depend  crucially on various assumptions regarding couplings with the inflaton and/or other fields, and on thermalization processes in an expanding cosmology; see the review\cite{reheat} for further references. The question of how the nearly $\simeq 100$ degrees of freedom of the Standard Model attain a state of local thermodynamic equilibrium after inflation and on what time scales is still unanswered.   Most studies \emph{model} the couplings and dynamics; therefore  any model of reheating is at best tentative and very approximate.  We bypass the inherent ambiguities and model dependence of the reheating dynamics, and  assume \emph{instantaneous} reheating after inflation  to a radiation dominated (RD) era.  The physical reason behind this assumption is that we  are primarily concerned with wavevectors that have crossed the Hubble radius during inflation well before the transition to (RD) and are well outside the horizon during this transition; hence they are causally decoupled from the microphysics of reheating. These modes feature very slow dynamics at the end of inflation, and the assumption that they are nearly frozen during the reheating time interval seems physically warranted (see further discussion in section (\ref{sec:discussion})).  We assume that both the scale factor and the Hubble rate are \emph{continuous} across the transition. Along with the continuity of fermion wave functions    across the transition, this, in fact, entails the continuity of the energy   momentum tensor. These aspects will be discussed in detail below.

 \textbf{iv:)} Unlike previous studies that invoked the adiabatic approximation, we study \emph{non-adiabatic} cosmological production. This is a direct consequence of a very small mass compared to the Hubble scale during inflation and  field fluctuations with superhorizon wavelengths after inflation.

 \textbf{v:)} Inflation is generically driven by a scalar field whose expectation value dominates the energy momentum tensor that sources gravity. The (RD) era is dominated by a large number $\gtrsim 100$ of ultrarelativistic degrees of freedom, therefore neglecting the back reaction of the (DM) degree of freedom is justified. Therefore, we   take the space time metric  during these eras as a \emph{background}.

In comoving
coordinates, the action is given by
\be
S   =  \int d^3x \; dt \;  \sqrt{-g} \,  \overline{\Psi}  \Big[i\,\gamma^\mu \;  \mathcal{D}_\mu -m  \Big]\Psi     \,. \label{lagrads}
\ee

 For Majorana fermions the action is multiplied by a factor $1/2$ (see appendix (\ref{app:majorana})).  Introducing  the vierbein field $e^\mu_a(x)$  defined as
$$
g^{\mu\,\nu}(x) =e^\mu_a (x)\;  e^\nu_b(x) \;  \eta^{a b} \; ,
$$
\noindent where $\eta_{a b}= \mathrm{diag}(1,-1,-1,-1)$ is the Minkowski space-time metric,
the curved space time Dirac gamma- matrices $\gamma^\mu(x)$  are given
by
\be
\gamma^\mu(x) = \gamma^a e^\mu_a(x) \quad , \quad
\{\gamma^\mu(x),\gamma^\nu(x)\}=2 \; g^{\mu \nu}(x)  \; ,
\label{gamamtx}\ee where the $\gamma^a$ are the Minkowski space time Dirac matrices, chosen to be in the standard Dirac representation. The fermion covariant derivative $\mathcal{D}_\mu$ is given in terms of the spin connection by\cite{weinbergbook,casta,parkerbook,birrell}

\be
\mathcal{D}_\mu   =    \partial_\mu + \frac{1}{8} \;
[\gamma^c,\gamma^d] \;  e^\nu_c  \; \left(\partial_\mu e_{d \nu} -\Gamma^\lambda_{\mu
\nu} \;  e_{d \lambda} \right) \,,  \label{fermicovader}
\ee  where $\Gamma^\lambda_{\mu
\nu}$ are the usual Christoffel symbols.

For a spatially flat Friedmann-Robertson-Walker cosmology
  in conformal time $d\eta= dt/a(t)$,   the metric becomes
\be
g_{\mu\nu}= a^2(\eta) \;  \eta_{\mu\nu}
 \,,\label{gmunu}
\ee

\noindent  and the vierbeins $e^\mu_a$ are given by
\be
 e^\mu_a = a^{-1}(\eta)\; \delta^\mu_a ~~;~~ e^a_\mu = a(\eta) \; \delta^a_\mu \,. \label{vierconf}\ee

The fermionic part of the action in conformal coordinates now becomes
\be S_f = \int d^3 x\,d\eta\,a^4(\eta)\,  \overline{\Psi}(\vec{x},\eta)\,\Bigg[ i \frac{\gamma^0}{a(\eta)}\,\Big(\frac{d}{d\eta} + 3 \frac{a^{'}(\eta)}{2a(\eta)}\Big) + i \,\frac{\gamma^i}{a(\eta)}\nabla_i - m   \Bigg]\Psi(\vec{x},\eta)\,. \label{Sf}\ee

The Dirac Lagrangian density in conformal time simplifies to
\be
\sqrt{-g} \; \overline{\Psi}\Big(i \; \gamma^\mu \;  \mathcal{D}_\mu
\Psi -m  \Big)\Psi  =
\big(a^{3/2}(\eta)\,\overline{\Psi}(\vec{x},\eta)\big) \;  \Big[i \;
{\not\!{\partial}}-m \; a(\eta) \Big]
\big(a^{3/2}(\eta)\,{\Psi}(\vec{x},\eta)\big)\,, \label{confferscal}
\ee
\noindent where $i {\not\!{\partial}}=\gamma^a \partial_a$ is the usual Dirac
differential operator in Minkowski space-time in terms of flat
space time $\gamma^a$ matrices. Introducing the conformally rescaled fields
\be  a^{\frac{3}{2}}(\eta)\,{\Psi(\vx,t)}= \psi(\vx,\eta)\,, \label{rescaledfields}\ee
  the action becomes
   \be  S    =
  \int d^3x \; d\eta \,   \overline{\psi} \;  \Big[i \;
{\not\!{\partial}}- M(\eta)   \Big]
 {\psi}    \;, \label{rescalagds}\ee
  with
   \be M (\eta) = m  \,a(\eta) \,.  \label{masfer}\ee

 The Dirac equation for the conformally rescaled fermi field becomes
\be  \Big[i \;
{\not\!{\partial}}- M(\eta)    \Big]
 {\psi}  = 0\,,    \label{diraceqn}\ee   and expand $ \psi({\vec x},\eta) $  in a comoving volume $V$  as
\be
\psi(\vec{x},\eta) =    \frac{1}{\sqrt{V}}
\sum_{\vec{k},s}\,   \left[b_{\vec{k},s}\, U_{s}(\vec{k},\eta) +
d^{\dagger}_{-\vec{k},s}\, V_{s}(-\vec{k},\eta)
 \right]\,e^{i \vec{k}\cdot\vec{x}} \; ,
\label{psiex}
\ee
  and  the spinor mode functions $U,V$ obey the  Dirac equations
\bea
&& \Bigg[i \; \gamma^0 \;  \partial_\eta - \vec{\gamma}\cdot \vec{k}
-M(\eta) \Bigg]U_s(\vec{k},\eta)   =  0 \label{Uspinor} \\
&& \Bigg[i \; \gamma^0 \;  \partial_\eta - \vec{\gamma}\cdot \vec{k} -M(\eta)
\Bigg]V_s(-\vec{k},\eta)   =   0 \,. \label{Vspinor}
\eea
 Multiplying the above equations both by $\gamma^0$ we find that
 \be \frac{d}{d\eta}\Big( U^\dagger_{s}(\vec{k},\eta)\,U_{s}(\vec{k},\eta)\Big) = 0 ~~;~~
 \frac{d}{d\eta}\Big( V^\dagger_{s}(-\vec{k},\eta)\,V_{s}(-\vec{k},\eta)\Big)=0~~;~~ \frac{d}{d\eta}\Big( U^\dagger_{s}(\vec{k},\eta)\,V_{s}(-\vec{k},\eta)\Big) =0 \,. \label{consta}\ee
We choose to work with the standard Dirac representation of the (Minkowski) $\gamma^a$ matrices.

It proves
convenient to write
\bea
U_s(\vec{k},\eta) & = & \Bigg[i \; \gamma^0 \;  \partial_\eta -
\vec{\gamma}\cdot \vec{k} +M(\eta)
\Bigg]f_k(\eta)\, u_s \label{Us}\\
V_s(-\vec{k},\eta) & = & \Bigg[i \; \gamma^0 \;  \partial_\eta -
\vec{\gamma}\cdot \vec{k} +M( \eta)
\Bigg]g_k(\eta)\,v_s \label{Vs}
\eea
\noindent with $u_s\,,\,v_s$ being
constant spinors  obeying
\be
\gamma^0 \; u_s  =  u_s
\label{Up} \qquad , \qquad
\gamma^0 \;  v_s  =  -v_s \,.
\ee

We choose the spinors $u_s \,;\,v_s$  as
\be u_s = \left(\begin{array}{c}
                             \xi_s\\
                            0
                          \end{array}\right)~~;~~ v_s = \left(\begin{array}{c}
                            0\\
                            \xi_s
                          \end{array}\right) \label{helispinors}\,, \ee  where the two component spinors $\xi_s$ are chosen to be helicity eigenstates, namely                         \be \vec{\sigma}\cdot\vec{k} = s \, k \, \xi_s ~~;~~ s = \pm 1\,.  \label{helicity} \ee
Inserting the ansatz (\ref{Us},\ref{Vs}) into the Dirac equations (\ref{Uspinor},\ref{Vspinor}) we find that the mode functions $f_k(\eta);g_k(\eta)$ obey the following
equations of motion
\bea \left[\frac{d^2}{d\eta^2} +
k^2+M^2 (\eta)-i \; M' (\eta)\right]f_k(\eta) & = & 0 \,, \label{feq}\\
\left[\frac{d^2}{d\eta^2} + k^2+M^2 (\eta)+i \; M' (\eta)\right]g_k(\eta)
& = & 0 \,,\label{geq}
\eea  where primes stand for  derivatives with respect to $\eta$.

We will adopt ``in''  boundary conditions for wave vectors deep inside the Hubble radius during inflation, so that as $-k\eta \rightarrow \infty$
\be f_k(\eta) ~\rightarrow ~e^{-ik\eta} ~~;~~ g_k(\eta) ~\rightarrow ~ e^{ik\eta}  \,.  \label{inbcs}\ee    With these boundary conditions, it follows from equations (\ref{feq},\ref{geq}) that
\be g_k(\eta) = f^*_k(\eta)\,.  \label{inrelation}\ee

Finally, the spinor solutions with ``in'' boundary conditions are

\be U_s(\vec{k},\eta) = N\,\left( \begin{array}{c}
                                     \mathcal{F}_k(\eta)\, \xi_s\\
                                    k\,f_k(\eta) \, s \, \xi_s
                                  \end{array}\right)\,,  \label{Uspin}\ee

 \be V_s(-\vec{k},\eta) = N\,\left( \begin{array}{c}
                                      -k\,f^*_k(\eta) \, s\,\xi_s\\
                                    \mathcal{F}^*_k(\eta)  \,   \xi_s
                                  \end{array}\right)\,,  \label{Vspin}\ee  where we introduced

 \be \mathcal{F}_k(\eta) =    if'_k(\eta)+M(\eta) f_k(\eta)\,, \label{capF}\ee and
                                  $N$ is a (constant) normalization factor.

                                   The spinor solutions are normalized as follows
 \be    U^\dagger_s(\vec{k},\eta)\,  U_{s'}(\vec{k},\eta) =\delta_{s,s'}~~;~~  V^\dagger_s(-\vec{k},\eta)\,  V_{s'}(-\vec{k},\eta) =\delta_{s,s'}\,,\label{normas}\ee yielding
 \be |N|^2\Big[\mathcal{F}^*_k(\eta)\,\mathcal{F}_k(\eta)+ k^2 f^*_k(\eta)\, f_k(\eta) \Big] = 1\,. \label{normaN} \ee With these normalization conditions the operators $b_{\vec{k},s},d_{\vec{k},s}$ in the field expansion (\ref{psiex})  obey the usual canonical anticommutation relations.

 Furthermore, it is straightforward to confirm that
\be U^\dagger_s(\vec{k},\eta)\,  V_{s'}(-\vec{k},\eta) =0 \,. \label{ortho}\ee

The spinors $U_s,V_s$ furnish a complete set of four   independent solutions of the Dirac equation.

The inflationary stage is  described by a   spatially flat de Sitter space time (thereby neglecting slow roll corrections)  with a scale factor
\be a(\eta) = -\frac{1}{H_{dS}(\eta-2\eta_R)} \,,\label{adS} \ee where $H_{dS}$ is the Hubble constant during de Sitter and $\eta_R$ is the (conformal) time at which the de Sitter stage transitions to the (RD) stage.

During the     (RD)  stage
\be H(\eta) = \frac{1}{a^2(\eta)}\frac{d a(\eta)}{d\eta} = 1.66 \sqrt{g}\,\frac{T^2_0}{M_{Pl}\,a^2(\eta)}\,, \label{hrd}\ee where $g$ is the effective number of ultrarelativistic degrees of freedom, which varies in time as different particles become non-relativistic. We take $g=2$ corresponding to radiation today. As discussed in detail in section (\ref{sec:discussion}) by taking $g=2$  we   obtain a \emph{lower bound} on the (DM) abundance and equation of state,  differing  by a factor of $\mathcal{O}(1)$ from the   abundance if the (RD) era is dominated only by standard model degrees of freedom.

 With this approximation the scale factor is given by
\be a(\eta)=  H_R\,\eta \,,  \label{Crdmd}\ee with
\be H_R= H_0\,\sqrt{\Omega_R}\simeq 10^{-35}\,\mathrm{eV}   \,,  \label{Hs}\ee and matter radiation equality occurs at
\be a_{eq}= \frac{\Omega_R}{\Omega_M} \simeq 1.66\,\times 10^{-4}  \,.\label{rands}\ee

The result (\ref{Hs}) corresponds to  the value of the fraction density $\Omega_R$ \emph{today}, thereby neglecting the change in the number of degrees of freedom contributing to the radiation density fraction.   If there are $g$ effective ultrarelativistic degrees of freedom, eqn. (\ref{Hs}) must be multiplied by $\sqrt{g/2}$. However, as discussed in detail in section (\ref{sec:discussion}) accounting for ultrarelativistic degrees of freedom of the standard model at the time of the transition between inflation and (RD)  modifies the final abundance by a factor of $\mathcal{O}(1)$.

We model the transition from de Sitter to (RD) at a (conformal) time $\eta_R$  by requiring that the scale factor and the Hubble rate be continuous across the transition at   $\eta_R$,   assuming self-consistently  that the transition occurs deep in the (RD) era so that $a(\eta_R) = H_R\,\eta_R \ll a_{eq}$.

 Using   $H(\eta) = a'(\eta)/a^2(\eta)$,  continuity of the scale factor and Hubble rate at $\eta_R$  imply that
\be a_{dS}(\eta_R) = \frac{1}{H_{dS}\,\eta_R}= H_R\,\eta_R ~~;~~ H_{dS} = \frac{1}{H_R\,\eta^2_R} \,, \label{transition} \ee yielding
\be \eta_R = \frac{1}{\sqrt{H_{dS}\,H_R}}\,.  \label{etaR}\ee

The most recent constraints from Planck\cite{planck2018} on the tensor-to-scalar ratio yields

\be H_{dS}/M_{Pl} < 2.5\times 10^{-5} ~~(95\%)\,\mathrm{CL} \,. \label{planckcons}\ee We take as a representative value $H_{dS} = 10^{13}\,\mathrm{GeV}$, from which it follows that
\be a_{dS}(\eta_R) =H_R\,\eta_R = \sqrt{\frac{H_R}{H_{dS}}} \simeq 10^{-28}\ll a_{eq} \,. \label{scalefac}\ee This scale corresponds to an approximate ambient radiation temperature after the transition from de Sitter to (RD)
\be T(\eta_R) \simeq \frac{T_0}{a_{RD}(\eta_R)} \simeq 10^{15}\,\mathrm{GeV} \label{Tgut}\ee where $T_0\propto 10^{-4}\,\mathrm{eV}$ is the CMB temperature today.

We focus on the case when the fermion is ``light'' as compared to the scale of inflation, namely $m \ll H_{dS}$, but otherwise arbitrary,   and  introduce the dimensionless ratio
\be \varepsilon = \sqrt{\frac{m}{H_{dS}}} \ll 1 \,,\label{eps} \ee  which will play an important role in the analysis.

\subsection{Matching conditions:}\label{subsec:match}

Defining $\psi^<(\vec{x},\eta)$ and $\psi^>(\vec{x},\eta)$ the fermion field for $\eta < \eta_R$ and
$\eta > \eta_R$ respectively, and because the Dirac equation (\ref{diraceqn}) is first order in time the Dirac field is continuous across the transition, the matching condition   is
\be \psi^<(\vec{x},\eta_R) = \psi^>(\vec{x},\eta_R)\,.\label{match} \ee
This continuity condition along with the continuity of the scale factor and Hubble rate at $\eta_R$  results in that the energy density, namely the expectation value of  $T^{0}_0$ is \emph{continuous at the transition}. This important aspect is discussed further in section (\ref{sec:tmunu}).

Introducing the Dirac spinors during the inflationary ($\eta < \eta_R$)  and radiation-dominated ($\eta > \eta_R$) dominated stages as $U^<\,,\,V^<$  and   $U^>\,,\,V^>$ respectively,  it follows from the matching condition (\ref{match}) that
\bea &&  U^<_{s}(\vec{k},\eta_R) = U^>_{s}(\vec{k},\eta_R)\,, \label{Umach}\\
&&  V^<_{s}(-\vec{k},\eta_R) = V^>_{s}(-\vec{k},\eta_R)\,. \label{Vmach}\eea

\subsection{Adiabatic vs. non-adiabatic evolution, asymptotic ``out'' particle states.}\label{subsec:adia}

Our goal is to solve  {exactly} the mode equations during the inflationary and (RD) stages and implement the matching conditions (\ref{match},\ref{Umach},\ref{Vmach}). During   inflation the mode equations are solved with the ``in'' boundary conditions (\ref{inbcs}) corresponding to the fermi fields being in the Bunch-Davies vacuum state (see next section). We now need to determine the boundary conditions on the mode functions during (RD).

Let us consider solving the mode equation (\ref{feq}) in a Wentzel-Kramers-Brillouin (WKB) adiabatic expansion, writing
\be f_k(\eta) = e^{-i \int^\eta \Omega_k(\eta')\,d\eta'} \,, \label{wkb}\ee we find that $\Omega_k(\eta)$ obeys
\be \Omega^2_k(\eta) + i\,\Omega'_k(\eta)-\omega^2_k(\eta) + i M'(\eta) = 0 ~~;~~ \omega^2_k(\eta) = k^2 + M^2(\eta)\,. \label{Omeq}\ee  Expanding $\Omega_k(\eta) = \Omega^{(0)}_k(\eta) + \Omega^{(1)}_k(\eta) +\cdots $ where the superscript implies order in a derivative adiabatic expansion, we find up to first order (see Appendix (\ref{app:ferad}))
\be \Omega_k(\eta) = \omega_k(\eta) \,\Big[1-  i \frac{\omega'_k(\eta)}{2\,\omega^2_k(\eta)}- i \frac{M'(\eta)}{2\,\omega^2_k(\eta)} +\cdots \Big]\,,  \label{wkbex}\ee where the dots stand for terms with higher order derivatives with respect to $\eta$. We refer to terms with $n$-derivatives as $n-th$ order adiabatic. We note that the term $M'(\eta)$ in the mode equation (\ref{feq}) is formally first order adiabatic, as is manifest in eqn. (\ref{wkbex}). This adiabatic expansion is reliable and useful provided that  terms of higher adiabatic order  are smaller order by order. To assess the reliability, consider the first order corrections displayed in (\ref{wkbex}), and writing them as follows
\be \frac{M'(\eta)}{ \omega^2_k(\eta)} = \frac{H(\eta)}{m}\,\frac{1}{\gamma^2_k(\eta)} ~~;~~ \frac{\omega'_k(\eta)}{ \omega^2_k(\eta)} = \frac{H(\eta)}{m}\,\frac{1}{\gamma^3_k(\eta)}\,,\label{firstordad}\ee where $H(\eta),\gamma_k(\eta)$ are the Hubble expansion rate and local Lorentz factor respectively, namely
\be H(\eta) = \frac{a'(\eta)}{a^2(\eta)}~~;~~ \gamma_k(\eta) = \Bigg[1+ \Big(\frac{k}{m\,a(\eta)}\Big)^2 \Bigg]^{1/2}\,.  \label{Handga}\ee During inflation and for superhorizon wavelengths, it follows that
\be \frac{M'(\eta)}{ \omega^2_k(\eta)}\simeq \frac{\omega'_k(\eta)}{ \omega^2_k(\eta)} \simeq \frac{H_{dS}}{m} \simeq \frac{1}{\varepsilon} \gg 1\,. \label{infladi}\ee Therefore, the adiabatic approximation  {fails} during the inflationary stage for superhorizon wavelengths and $m \ll H_{dS}$.

During the (RD) era and for very long-wavelength modes,

\be \frac{M'(\eta)}{ \omega^2_k(\eta)}\simeq \frac{\omega'_k(\eta)}{ \omega^2_k(\eta)} \simeq \frac{H_R}{m\,a^2(\eta)} \,, \label{RDdi}\ee therefore the adiabatic expansion becomes reliable for
\be a(\eta) \gg \frac{10^{-17}}{\sqrt{m/(\mathrm{eV})}}\,. \label{Radadia}\ee Even for $m$ as small as $10^{-22}\,\mathrm{eV}$ the adiabatic expansion becomes reliable prior to matter radiation equality. We anticipate that the most interesting range for fermionic (DM) is $m \gg \mathrm{GeV}$ (see section (\ref{sec:tmunu}) below), hence the adiabatic approximation becomes very reliable for $a(\eta) \gg 10^{-22}$.  Two important points follow from this analysis: \textbf{i:)} during inflation and in the early stages of (RD) following the transition from inflation, the adiabatic approximation is not reliable in the range $10^{-28} \lesssim a(\eta) \lesssim 10^{-22}$, \textbf{ii:)}   {near} matter radiation equality ($a_{eq} \simeq 10^{-4}$) the adiabatic approximation to  {zeroth} order is very reliable. Therefore, the mode functions both during inflation and the early stages after the transition to  (RD) must be found  {exactly},  and the asymptotic ``out'' boundary conditions for these modes during (RD) can be reliably defined in the asymptotic adiabatic regime. Appendix (\ref{app:ferad}) provides more technical details on the nature of the adiabatic expansion for Fermi fields.

In summary: we do  {not} invoke the adiabatic approximation during inflation or the early stages after the transition to (RD), solving exactly for the mode functions during these stages. However, we {do} invoke it to determine the asymptotic ``out'' boundary conditions on the mode functions and spinors during the (RD) era. We refer to the solutions of eqn. (\ref{feq}) for the mode functions $f_k(\eta)$ during the (RD) era obeying the asymptotic ``out'' boundary conditions just prior to $a_{eq}$, as asymptotic ``out''  {particle} states.
\be f_k(\eta) ~~\rightarrow ~~ e^{-i \int^\eta \omega_k(\eta')\,d\eta'} = e^{-i\int^t E_k(t')\, dt'}\,,\label{asyfok}\ee where in comoving time
\be E_k(t) = \sqrt{k^2_{ph}(t)+m^2} ~~;~~ k_{ph}(t)= k/a(\eta(t))\,.  \label{energy}\ee

In the next section we will solve    for the mode functions $f_k(\eta)$ and the spinor solutions of the Dirac equation both during inflation and radiation domination with the ``in'' and ``out'' (particle) asymptotic boundary conditions,
\bea  && f_k(\eta) ~~ {}_{\overrightarrow{-k\eta \rightarrow \infty} } ~~ e^{-ik\eta} ~~\mathrm{IN~~(inflation)}\,,\label{ininfla} \\ && f_k(\eta) ~~ {}_{\overrightarrow{a(\eta) \simeq a_{eq}} } ~~ e^{-i\int^\eta \omega_k(\eta')\,d\eta'} ~~\mathrm{OUT~~(RD)}\,, \label{outRD}\eea identifying the solutions with this ``out'' boundary conditions as describing particle states. The mode functions $g_k(\eta) = f^*_k(\eta)$ and the corresponding spinors  are associated with the anti-particle ``out'' states. We then match the respective solutions at $\eta=\eta_R$ via the matching conditions (\ref{Umach},\ref{Vmach}).

\section{Exact solutions:}\label{sec:solutions}

\subsection{Inflationary stage:}\label{subsec:infla}

We   consider that the inflationary stage is described by an exact de Sitter space time with scale factor given by eqn. (\ref{adS}) and that the fermionic degrees of freedom are in the Bunch-Davies vacuum state during inflation. This implies that consistently with (\ref{inbcs})   the solutions $U_s(\vec{k},\eta); V_s(-\vec{k},\eta)$ obey the ``in'' boundary conditions

 \be U_s(\vec{k},\eta) \rightarrow e^{-ik\eta}~;~ V_s(\vec{k},\eta) \rightarrow e^{ik\eta}\,, \label{inbc}\ee  for wavevectors deep inside the Hubble radius $-k\eta \rightarrow \infty$, given by (\ref{Uspin},\ref{Vspin})  along with
 \be b_{\vec{k},s}|0\rangle =0 ~~;~~  d_{-\vec{k},s}|0\rangle =0 \,, \label{bdbc}\ee

  The equations (\ref{feq},\ref{geq}) for the mode functions becomes
\be \Bigg[ \frac{d^2}{d\tau^2} + k^2 - \frac{\nu^2-1/4}{\tau^2}\Bigg]f_k(\tau) = 0 ~~;~~ \tau = \eta-\eta_R~~;~~ \nu = \frac{1}{2}-i\,\varepsilon^2 \,. \label{desieqn} \ee in terms of the dimensionless ratio (\ref{eps}). The solution with ``in'' boundary conditions (\ref{ininfla}), namely $f_k(\eta) \rightarrow e^{-ik\eta} $ for sub-Hubble modes, is given by
\be f_k(\tau) = \sqrt{-\frac{\pi k \tau}{2}}~e^{i\pi(\nu+1/2)/2}~H^{(1)}_\nu(-k\tau) ~~;~~ g_k(\tau) = f^*_k(\tau)\,, \label{fgdS}\ee where $H^{(1)}_{\nu}$ is a Hankel function.

Therefore with Bunch-Davies boundary conditions, the spinors $U^<(\vec{k},\eta)~;~V^<(-\vec{k}\eta)$ are given by the expressions (\ref{Uspin},\ref{Vspin}) with $f_k(\eta)$ given by (\ref{fgdS}). In the superhorizon limit   the solution (\ref{fgdS}) behaves as
\be f_k(\tau) \propto \big[-k\tau\big]^{i\varepsilon^2}\,, \label{shlim}\ee therefore unlike the case of bosonic degrees of freedom, these mode functions are  {not} enhanced for superhorizon wavelengths with the modulus remaining $\mathcal{O}(1)$. Furthermore, as it will become clear below,  the relevant  dimensionless comoving momentum is (see below) $q = k/\sqrt{m H_R}$ from which  it follows that at the end of inflation $k\eta_R = q \,\varepsilon \ll 1$. It will be shown below (see section (\ref{sec:tmunu})) that the abundance of produced (DM) particles is dominated by the region $q \simeq 1$ hence the phase in (\ref{shlim}) is very slowly varying at the end of inflation for the relevant modes.

\subsection{Radiation dominated stage:}\label{subsec:RD}
We define the mode functions during (RD) as $h_k(\eta)$ to distinguish them from the solutions during the inflationary era. These obey the mode equations
\be \Bigg[ \frac{d^2}{d\eta^2}+\omega^2_k(\eta)   - i\,m H_R  \Bigg]h_k(\eta) = 0 ~~;~~\omega^2_k(\eta)= k^2 + m^2 H^2_R \eta^2 \,.\label{heqn} \ee

During the (RD) stage we introduce the spinors $\mathcal{U},\mathcal{V}$    that describe asymptotic   particle and anti-particle ``out'' states at long time  respectively. These are solutions of the Dirac equation during the (RD) era  satisfying the  asymptotic ``out''  boundary conditions (\ref{outRD}) yielding
\be  \mathcal{U}(\vec{k},\eta) \rightarrow \propto \,  e^{-i\int^\eta \omega_k(\eta') \,d\eta'}~~;~~ \mathcal{V}(\vec{k},\eta) \rightarrow \propto \, e^{i\int^\eta \omega_k(\eta') \,d\eta'}\,.  \label{rdbcs}\ee   This ``out'' boundary condition corresponds to the mode functions $h_k(\eta)$ solutions of (\ref{heqn})
 with the asymptotic behavior (\ref{outRD}), namely
\be h_k(\eta) \rightarrow  e^{-i\int^\eta \omega_k(\eta') \,d\eta'} \,. \label{hout}\ee

With these boundary conditions we find that these particle-antiparticle spinors are given by
\be  \mathcal{U}_s(\vec{k},\eta) = \widetilde{N}\,\left( \begin{array}{c}
                                     \mathcal{H}_k(\eta)\, \xi_s\\
                                    k\,h_k(\eta) \, s \, \xi_s
                                  \end{array}\right)\,,  \label{Uspinrd}\ee

 \be \mathcal{V}_s(-\vec{k},\eta) = \widetilde{N}\,\left( \begin{array}{c}
                                      -k\,h^*_k(\eta) \, s\,\xi_s\\
                                    \mathcal{H}^*_k(\eta)  \,   \xi_s
                                  \end{array}\right)\,,  \label{Vspinrd}\ee  where we have introduced
\be \mathcal{H}_k(\eta) =  ih'_k(\eta)+M(\eta) h_k(\eta)\,,   \label{calH}\ee   and
                                  $\widetilde{N}$ is a (constant) normalization factor chosen  so that
 \be     \mathcal{U}^\dagger_s(\vec{k},\eta)\,  \mathcal{U}_{s'}(\vec{k},\eta) =\delta_{s,s'}~~;~~  \mathcal{V}^\dagger_s(-\vec{k},\eta)\,  \mathcal{V}_{s'}(-\vec{k},\eta) =\delta_{s,s'}\,,\label{normasrd}\ee  yielding
 \be | \widetilde{N}|^2 \Big[ \mathcal{H}^*_k(\eta) \mathcal{H}_k(\eta)+ k^2 h^*_k(\eta)h_k(\eta) \Big] =1\,.  \label{nortilN}\ee
Again,  it is straightforward to confirm that
\be \mathcal{U}^\dagger_s(\vec{k},\eta)\,  \mathcal{V}_{s'}(-\vec{k},\eta) =0 \,. \label{orthord}\ee

These form a complete set of four solutions of the Dirac equation ($s=\pm 1$) during (RD).

It is convenient to introduce the following dimensionless combinations,
\be z= \sqrt{m H_R}\,\, \eta ~~;~~ q = \frac{k}{\sqrt{m H_R}} ~~;~~ \lambda = q^2 - i \label{dimcombos}\ee in terms of which eqn. (\ref{heqn}) becomes
\be \frac{d^2}{dz^2} h_k(z) + (z^2 + \lambda)h_k(z) =0 \,, \label{dimeqn}\ee the solutions of which are the parabolic cylinder functions\cite{gr,as,nist,bateman,magnus}
\be D_\alpha[\sqrt{2} e^{i\pi/4} z]~~;~~ D_\alpha[\sqrt{2} e^{3i\pi/4} z]~~;~~\alpha = -\frac{1}{2}-i \frac{\lambda}{2} = -1 - i\,\frac{q^2}{2}\,.  \label{solus}\ee The solution that fulfills the ``out'' boundary condition (\ref{hout}) (see appendix A) is given by
\be h_k(\eta) =  D_\alpha[\sqrt{2} e^{i\pi/4} z] \,. \label{soluh}\ee The general solution for the spinor wave functions $U^>,V^>$ during the (RD) era are linear combinations of the four independent solutions (\ref{Uspinrd},\ref{Vspinrd}). In principle, with four independent solutions during inflation matching onto four independent solutions during (RD) there would be a $4\times 4$ matrix of Bogoliubov coefficients,  however,  because helicity is conserved, the linear combinations are given by
\bea &&  U^>_s(\vec{k},\eta) = A_{k,s}\, \mathcal{U}_s(\vec{k},\eta)+ B_{k,s} \,\mathcal{V}_s(-\vec{k},\eta)\label{Ugreat} \\
&& V^>_s(-\vec{k},\eta) = C_{k,s}\, \,\mathcal{V}_s(-\vec{k},\eta)+
D_{k,s}\, \mathcal{U}_s(\vec{k},\eta) \,.  \label{Vgreat} \eea The  Bogoliubov coefficients $A_{k,s} \cdots D_{k,s}$ are obtained from the matching conditions (\ref{Umach},\ref{Vmach}) and the relations (\ref{normasrd},\ref{orthord}). We find
\bea && A_{k,s} =  \mathcal{U}^\dagger_s(\vec{k},\eta_R) U^<_s(\vec{k},\eta_R) = N\widetilde{N} \Big[\mathcal{H}^*_k\,\mathcal{F}_k+k^2 h^*_k f_k \Big]_{\eta=\eta_R} \label{Ak}\\
&& B_{k,s} =  \mathcal{V}^\dagger_s(-\vec{k},\eta_R) U^<_s(\vec{k},\eta_R) = N\widetilde{N}\,s\, \Big[-k h_k \mathcal{F}_k +k f_k \mathcal{H}_k  \Big]_{\eta=\eta_R}\label{Bk} \\
&& C_{k,s} =  \mathcal{V}^\dagger_s(-\vec{k},\eta_R) V^<_s(-\vec{k},\eta_R) = N\widetilde{N} \Big[k^2 h_k f^*_k + \mathcal{H}_k\,\mathcal{F}^*_k  \Big]_{\eta=\eta_R}\label{Ck}\\
&& D_{k,s} =  \mathcal{U}^\dagger_s(\vec{k},\eta_R) V^<_s(-\vec{k},\eta_R) = N\widetilde{N} \,s\, \Big[ -kf^*_k \mathcal{H}^*_k + k h^*_k \mathcal{F}^*_k  \Big]_{\eta=\eta_R}\,. \label{Dk} \eea From these relations we find the important corollary
\be D_{k,s} = - B^*_{k,s} ~~;~~ C_{k,s} = A^*_{k,s} \,, \label{bogorels}\ee which guarantees the orthogonality
\be  {U^>_s}^\dagger(\vec{k},\eta) V^>_s(-\vec{k},\eta) =0 \,. \label{ortogreat}\ee
Furthermore the normalization conditions (\ref{normas},\ref{ortho},\ref{normasrd},\ref{orthord}) yield the following relations between Bogoliubov coefficients
\be |A_{k,s}|^2 + |B_{k,s}|^2 = |C_{k,s}|^2+ |D_{k,s}|^2 =1 \,, \label{bogoconst}\ee which can be confirmed straightforwardly from the expressions (\ref{Ak}-\ref{Dk}) and the normalization conditions.

During the (RD) era, with $U_s \equiv U^>_s; V_s \equiv V^>_s$ with $U^>, V^>$ given by (\ref{Ugreat},\ref{Vgreat}) the field expansion (\ref{psiex}) becomes

 \be
\psi(\vec{x},\eta) =    \frac{1}{\sqrt{V}}
\sum_{\vec{k},s}\,   \left[\widetilde{b}_{\vec{k},s}\, \mathcal{U}_{s}(\vec{k},\eta) +
\widetilde{d}^{\,\dagger}_{-\vec{k},s}\, \mathcal{V}_{s}(-\vec{k},\eta)
 \right]\,e^{i \vec{k}\cdot\vec{x}} \; ,
\label{psiexrd}
\ee where
\bea \widetilde{b}_{\vec{k},s} & = &   {b}_{\vec{k},s} A_k + {d}^{\dagger}_{-\vec{k},s} D_{k,s} \label{btil} \\
\widetilde{d}^{\,\dagger}_{-\vec{k},s} & = &   {d}^{\dagger}_{-\vec{k},s} C_{k,s} + {b}_{\vec{k},s} B_{k,s} \,. \label{ddtil}
\eea The relations (\ref{bogoconst}, \ref{bogorels}) entail that the new operators $\widetilde{b},\widetilde{d}$ obey the canonical anticommutation relations. The operators $\widetilde{b}$ and $\widetilde{d}$ create asymptotic particle and antiparticle states respectively.  In particular we find that the number of asymptotic ``out'' particle and antiparticle states in the Bunch-Davies vacuum state (\ref{bdbc}) are given by
\be  \langle 0| \widetilde{b}^\dagger_{\vec{k},s}   \widetilde{b}_{\vec{k},s}|0\rangle    =   |D_{k,s}|^2 = \langle 0| \widetilde{d}^\dagger_{-\vec{k},s}   \widetilde{d}_{-\vec{k},s}|0\rangle    =    |B_{k,s}|^2 \,. \label{parts}\ee We identify $|B_{k,s}|^2$ with the \emph{distribution function of produced particles}. The relation (\ref{bogoconst}) implies that
\be |B_{k,s}|^2 \leq 1\,, \label{maxn} \ee for each polarization $s$,  consistent with Pauli exclusion.

\subsection{Bogoliubov coefficients:}\label{subsec:bogos}

The Bogoliubov coefficients are obtained from the relations (\ref{Ak}-\ref{Dk}) where the scalar products of spinors are evaluated at the transition time $\eta=\eta_R$. For light fermions with $m/H_{dS} \ll 1$ ($\varepsilon \ll 1$)  these can be greatly simplified with the following approximations:

\textbf{i:)} During the inflationary era and  taking $\nu=1/2$ for $\varepsilon \ll 1$ in the solutions (\ref{fgdS})) yields
\be f_k(\eta_R) = e^{ik\eta_R} \simeq 1 \,,\label{fetaR}\ee where we have considered modes that are super-Hubble at the transition time, namely $|k\eta_R| \ll 1$. Furthermore, with
\be M(\eta_R) = \frac{m H_R}{\sqrt{H_R H_{dS}}} = \varepsilon \, \sqrt{m H_R}   \,, \label{MR} \ee
the normalization constant $N$ in the spinors (\ref{Uspin},\ref{Vspin}) is obtained by normalizing the spinors at $\eta=\eta_R$. In terms of the dimensionless ratio $q = k/\sqrt{m H_R}$ to lowest order in $\varepsilon$  at $\eta = \eta_R$ these spinors are given   by
\bea && U^<_s(\vec{k},\eta_R) = \frac{1}{\sqrt{2q(q+\varepsilon)}}\, \left( \begin{array}{c}
                                     (q+\varepsilon)\, \xi_s\\
                                    q \, s \, \xi_s
                                  \end{array}\right)\,  \label{UspiR}\\
&& V^<_s(-\vec{k},\eta_R) = \frac{1}{\sqrt{2q(q+\varepsilon)}}\,   \left( \begin{array}{c}
                                      q \, s\,\xi_s\\
                                    (q+\varepsilon)  \,   \xi_s
                                  \end{array}\right)\,.\label{VspiR}                               \eea

\textbf{ii:)} During radiation domination,   it follows from the definitions (\ref{dimcombos}) that at the transition time,
\be z_R = \sqrt{\frac{m H_R}{H_R H_{dS}}} = \varepsilon \ll 1 \,,\label{zetar}\ee and for $z \ll 1$ the parabolic cylinder functions feature an expansion in $z_R$ and $q \, z_R = k\eta_R \ll 1$ (see appendix (\ref{app:solutionRD})) for superhorizon wavevectors. Therefore we can safely approximate $z_R =0$ in the argument of the parabolic cylinder functions yielding the following identities,
\bea h_k(\eta_R) & = &   \frac{\sqrt{\pi/2}\,\,\,2^{-iq^2/4}}{\Gamma[1+i \frac{q^2}{4}]} \,,\label{ah}\\
h'_k(\eta_R) & = & -\frac{\sqrt{2\pi}\,e^{i\pi/4}2^{-iq^2/4}\,\sqrt{mH_R}}{\Gamma[\frac{1}{2}+i \frac{q^2}{4}]}\,. \label{ahpri} \eea

\textbf{iii:)} The result $|B_{k,s}|^2 \leq 1$ implies that there are no infrared divergences in the distribution function of particles, and in integrals,  the small $k$ region is suppressed by phase space. Furthermore, we anticipate, and prove below self-consistently, that in typical integrals involving $|B_{k,s}|^2$ the most relevant region is $k \simeq \sqrt{mH_R}$, namely $q\simeq 1$ (see below). Therefore, consistently with neglecting the $\mathcal{O}(\varepsilon)$ terms in (\ref{ah},\ref{ahpri})  we  set $\varepsilon \rightarrow 0$ in the spinors (\ref{UspiR},\ref{VspiR}).   With these results, the normalization constant $\widetilde{N}$ is obtained from the normalization conditions (\ref{normasrd}) at $\eta = \eta_R$, it is given by (see appendix (\ref{app:BogoBk}))
\be \widetilde{N} = \frac{e^{-\pi q^2/8}}{\sqrt{2mH_R}}\,. \label{normar}\ee

\begin{figure}[ht!]
\begin{center}
\includegraphics[height=5in,width=4.5in,keepaspectratio=true]{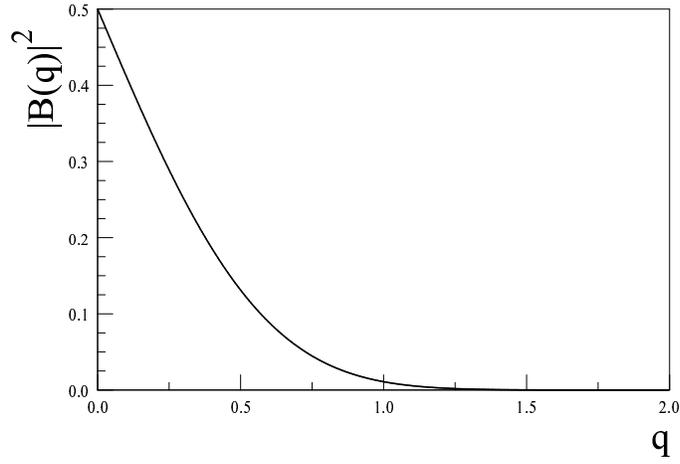}
\caption{The distribution function of produced particles $|B_{k,s}|^2\equiv |B(q)|^2$ vs $q= k/\sqrt{mH_R}$.  }
\label{fig:dist}
\end{center}
\end{figure}

\begin{figure}[ht!]
\begin{center}
\includegraphics[height=5in,width=4.5in,keepaspectratio=true]{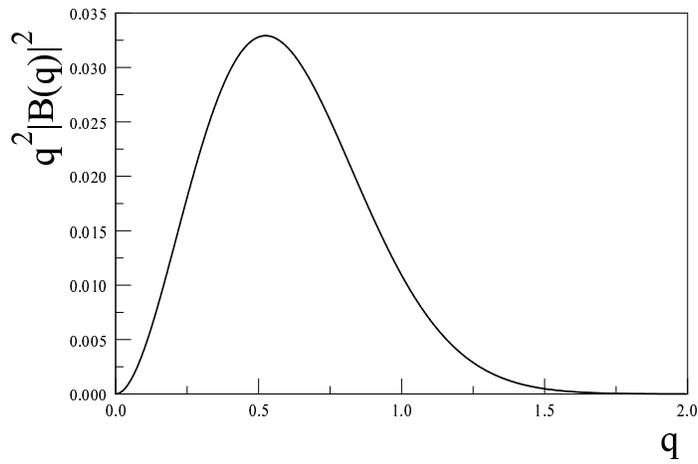}
\caption{The integrand for the total abundance $q^2|B_{k,s}|^2$ vs $q= k/\sqrt{mH_R}$. The abundance is dominated by typical momenta $k \simeq \sqrt{mH_R}$.   }
\label{fig:q2dist}
\end{center}
\end{figure}

With the approximations discussed above, and from the result (\ref{Bk}) we find
\be |B_{k,s}|^2 = \frac{1}{2} \,\Bigg[1- \widetilde{N} \, q\,\sqrt{mH_R}\,\Big(\mathcal{H}^*_k(\eta_R)\,h_k(\eta_R)+\mathcal{H}_k(\eta_R)\,h^*_k(\eta_R) \Big) \Bigg]\,. \label{bogofin1} \ee The calculation of the second term in the bracket is discussed in detail in appendix (\ref{app:BogoBk}) with the result,
\be |B_{k,s}|^2 = \frac{1}{2}\Bigg[1-\Big(1-e^{-\pi q^2} \Big)^{1/2}\ \Bigg]\,,\label{BogoBk} \ee yielding the behaviour
\be |B_{k,s}|^2 ~~{}_{\overrightarrow{k\rightarrow 0}}~~ \frac{1}{2}~~;~~|B_{k,s}|^2 ~~{}_{\overrightarrow{k \gg \sqrt{m H_R} }} ~~ \frac{1}{4}~e^{-\frac{2\pi\,k^2}{2 m H_R}}\equiv \frac{1}{4} |B_{mb}(k)|^2\,.\label{asybogoB}\ee The long-wavelength limit agrees with  ref.\cite{chungfer}. Remarkably, up to the prefactor $1/4$, for  $k \gtrsim \sqrt{mH_R}$ the Bogoliugov coefficient yields a Maxwell-Boltzmann distribution function ($|B_{mb}(k)|^2$) for a non-relativistic particle at an ``emergent'' temperature
\be T_H = \frac{H_R}{2\pi} \simeq 10^{-36}\,\mathrm{eV}\, , \label{Thawk}\ee and vanishing chemical potential in agreement with the result (\ref{parts}) which indicates that the number of produced particles equals that of anti-particles. Figs. (\ref{fig:dist}; \ref{fig:q2dist}) display  $|B(q)|^2$ and $q^2|B(q)|^2$ vs. $q$ respectively.

\vspace{1mm}

\begin{figure}[ht!]
\begin{center}
\includegraphics[height=5in,width=4.5in,keepaspectratio=true]{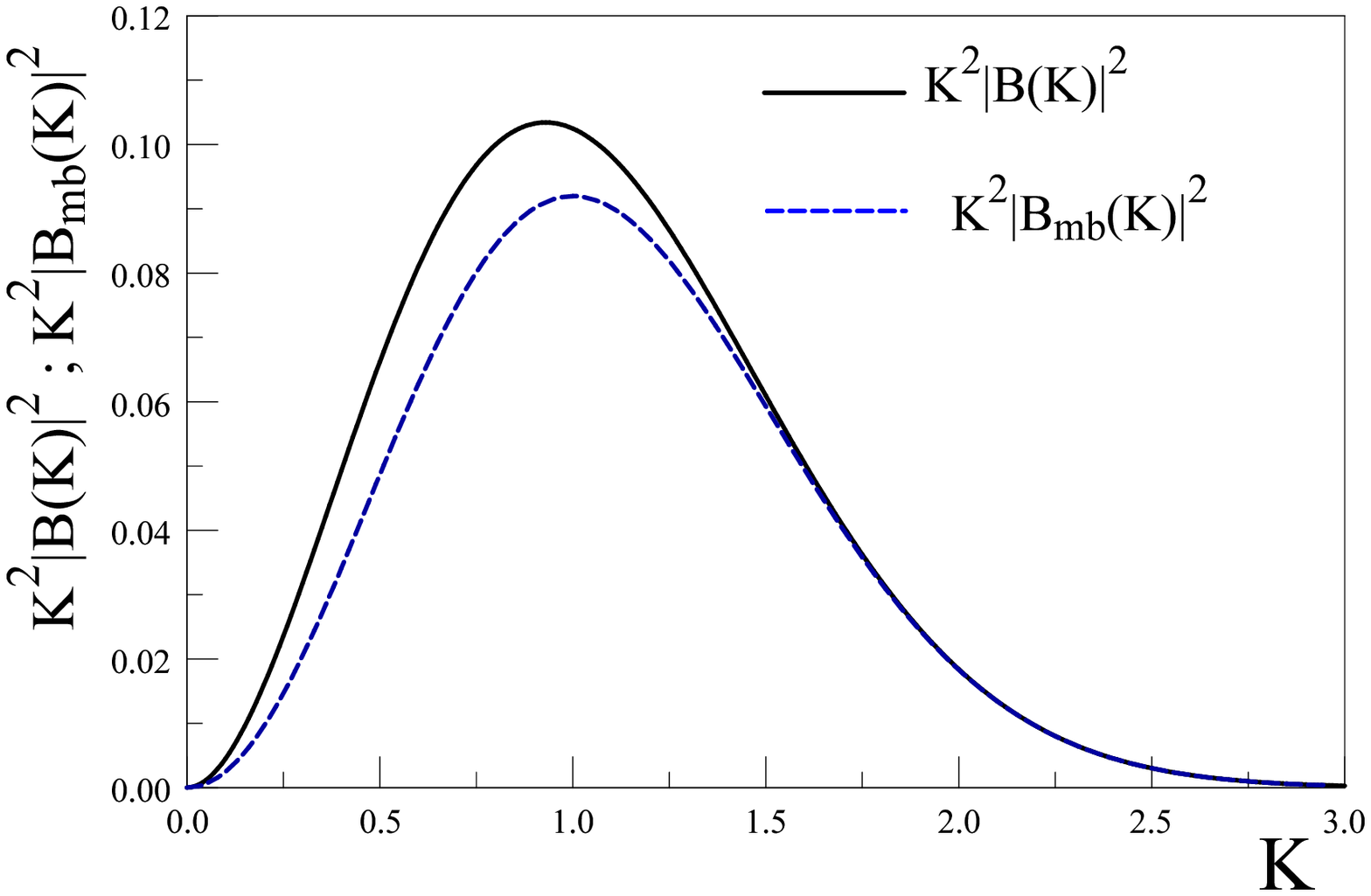}
\caption{The distributions  $K^2|B_{k,s}|^2$ and $\frac{K^2}{4}|B_{mb}(k)|^2 = \frac{K^2}{4}\,e^{-K^2}$  vs $K= k/\sqrt{2m T_H}$. The near agreement in the region of momenta $k\simeq \sqrt{2m T_H}$, which dominates the integrals,  entails a near thermal abundance.  }
\label{fig:comparison}
\end{center}
\end{figure}

Writing the  {distribution function} as
\be |B_{k,s}|^2 = \frac{1}{2} \Big[1-\big(1-e^{-\frac{k^2}{2mT_H}}\big)^{1/2}\Big]  \label{almMB}\ee makes manifest the striking similarity with a Maxwell-Boltzmann distribution function of a non-relativistic particle in thermal equilibrium at temperature $T_H$ and vanishing chemical potential, up to an overall normalization. Clearly the similarity does not hold for the longest wavelengths, which however are suppressed by the phase space factor,  but for $\frac{k^2}{2m} \gtrsim T_H$ the difference is small. Fig. (\ref{fig:comparison}) compares  $K^2 |B_{k,s}|^2 $  to $\frac{K^2}{4} |B_{mb}(k)|^2 \equiv \frac{K^2}{4}{e^{-K^2}}  $ vs. $K= \frac{k}{\sqrt{2mT_H}}$. The maximum difference is $\lesssim 10\,\%$, and  occurs at low momenta.

An important note is that the distribution function is localized in the region $q \simeq 1$, namely in the  range of momenta $ k \simeq \sqrt{m H_R}$, for which  $k\eta_R \lesssim \sqrt{\frac{mH_R}{H_{dS}H_R}}\lesssim \varepsilon \ll 1$. Therefore,  the largest contribution to the distribution function, hence the abundance and equation of state, is from wavelengths that are well outside the horizon at the end of inflation. This analysis confirms self-consistently the validity of one of the main assumptions, namely that of focusing on  superhorizon wavelengths at the end of inflation.

\section{Energy momentum tensor: renormalization  and asymptotics:}\label{sec:tmunu}

 \subsection{Energy density and pressure:}\label{subsec:EP}

 The energy momemtum tensor for Dirac fields is given by \cite{parkerbook}
 \be T^{\mu \nu} = \frac{i}{2} \Big( \overline{\Psi} \gamma^\mu  \stackrel{\leftrightarrow}{\mathcal{D}^\nu}   \,\Psi \Big) + \mu \leftrightarrow \nu \label{tmunudirac} \ee

 For Majorana fermions $T^{\mu \nu}$ must be multiplied by an extra factor $1/2$ (see appendix (\ref{app:majorana})). The expectation value of the energy momentum tensor in the Bunch-Davies vacuum state is given by
 \be \langle 0|T^\mu_\nu|0\rangle = \mathrm{diag}\big(\rho(\eta),-P(\eta),-P(\eta),-P(\eta)\big) \,. \label{EMT}\ee
In terms of conformal time and the conformally rescaled fields (\ref{rescaledfields})
 The energy density $\rho$  and pressure  $P$ are given by
 \be \rho(\eta) = \langle 0| T^0_0 |0\rangle= \frac{i}{ 2 a^4(\eta)} ~  \langle 0| \Big(\psi^\dagger (\vec{x},\eta) \frac{d}{d\eta}\,\psi (\vec{x},\eta) - \frac{d}{d\eta}\,\psi^\dagger (\vec{x},\eta) \,\psi(\vec{x},\eta) \Big)|0 \rangle  \,, \label{rho}\ee

 \be P(\eta) =  -\frac{1}{3} \sum_{j} \langle 0|T^j_j|0\rangle   = \frac{-i}{6 a^4(\eta)} ~ \langle 0| \Big(\psi^\dagger (\vec{x},\eta) \, \vec{\alpha}\cdot \vec{\nabla} \,\psi (\vec{x},\eta)-\vec{\nabla}\psi^\dagger (\vec{x},\eta)\cdot \vec{\alpha} \,\psi (\vec{x},\eta)\Big)|0 \rangle  \,, \label{P}\ee

 where $\vec{\alpha} = \gamma^0\,\vec{\gamma}$ and the expectation value is in the Bunch-Davies vacuum state. With the field expansion (\ref{psiex}) we find

 \be \rho(\eta) =  \frac{i}{2 a^4(\eta)} ~ \int^\infty_0  \sum_{s}\, \Bigg[V^\dagger_s(-\vec{k},\eta)\,\frac{d}{d\eta}V_s(-\vec{k},\eta)- \frac{d}{d\eta}{{V^\dagger}}_s(-\vec{k},\eta)\,V_s(-\vec{k},\eta) \Bigg]\, \frac{d^3k}{(2\pi)^3}\,, \label{rhoV}\ee

 \be P(\eta) =   \frac{1}{3\, a^4(\eta)} ~ \int^\infty_0  \sum_{s}\, V^\dagger_s(-\vec{k},\eta)\,\vec{\alpha}\cdot \vec{k} \,V_s(-\vec{k},\eta)~\frac{d^3k}{(2\pi)^3} \,. \label{PV}\ee

 Using the Dirac equation, it is straightforward to confirm covariant energy conservation, and also  that $\rho(\eta)$ can be written as
 \be \rho(\eta) =  \frac{1}{ a^4(\eta)} ~ \int^\infty_0  \sum_{s} V^\dagger_s(-\vec{k},\eta)\,\Big[ \vec{\alpha}\cdot \vec{k}+\gamma^0 M(\eta)\Big] \,V_s(-\vec{k},\eta) \, \frac{d^3k}{(2\pi)^3}\,,  \label{rho2}\ee yielding

 \be \langle 0|T^\mu_\mu|0\rangle =  \frac{m}{ a^3(\eta)} ~ \int^\infty_0  \sum_{s} \overline{V}_s(-\vec{k},\eta) \,V_s(-\vec{k},\eta) \,\frac{d^3k}{(2\pi)^3}\,.\label{trace}\ee

  Furthermore,  the continuity of the scale factor and Hubble rate and the continuity condition (\ref{match}) ensure that the energy momentum tensor is  {continuous} across the transition from inflation to radiation domination.

   During the inflationary stage the $V$ spinors are given by (\ref{Vspin}) with (\ref{fgdS}) and the energy momentum tensor yields the Bunch-Davies \emph{vacuum} energy density and pressure, which will be  fully subtracted in the renormalization of the energy momentum tensor (see below) .

 During the (RD) stage the spinors $V_s \equiv V^>_s$ are given by (\ref{Vgreat}) in terms of the Bogoliubov coefficients and the spinors $\mathcal{U},\mathcal{V}$ with particle and anti-particle ``out'' boundary conditions.
 Both the energy density and pressure feature \emph{three} distinct terms:
 \bea   \rho(\eta)  & = &  \rho_{vac}(\eta)+ \rho_{int}(\eta)+ \rho_{pp}(\eta)\label{3rhos}\\
  P(\eta)  & = &  P_{vac}(\eta)+ P_{int}(\eta)+ P_{pp}(\eta)\label{3Ps}\,, \eea where

\bea && \rho_{vac}    =      \frac{1}{  a^4(\eta)} ~ \int^\infty_0  \sum_{s=\pm 1}\, \Big[\mathcal{V}^\dagger_s(-\vec{k},\eta)\,\Sigma(\vec{k},\eta)\, \,\mathcal{V}_s(-\vec{k},\eta)  \Big]\, \frac{d^3k}{(2\pi)^3}\,,\label{rhovac} \\
&& \rho_{int}    =      -\frac{1}{ a^4(\eta)} ~ \int^\infty_0  \sum_{s=\pm 1}\,  \Big[A_{k,s}\,B^*_{k,s}  \,\mathcal{V}^\dagger_s(-\vec{k},\eta)\,\Sigma(\vec{k},\eta)\, \,\mathcal{U}_s(\vec{k},\eta) +h.c. \Big]\, \frac{d^3k}{(2\pi)^3}\,,\label{rhoint}\\
&&  \rho_{pp}   =       \frac{1}{  a^4(\eta)} ~ \int^\infty_0  \sum_{s=\pm 1}\, |B_{k,s}|^2 \Big[\mathcal{U}^\dagger_s(\vec{k},\eta)\,\Sigma(\vec{k},\eta)\, \,\mathcal{U}_s(\vec{k},\eta)
      -\mathcal{V}^\dagger_s(-\vec{k},\eta)\,\Sigma(\vec{k},\eta)\, \,\mathcal{V}_s(-\vec{k},\eta)  \Big]\, \frac{d^3k}{(2\pi)^3}\,,\nonumber \\\label{rhopp1}   \eea where
 \be \Sigma(\vec{k},\eta)=\vec{\alpha}\cdot \vec{k} + \gamma^0 M(\eta)\,,  \label{enesig}\ee is the conformal time instantaneous Dirac Hamiltonian, and

 \bea P_{vac} & = &   \frac{1}{3  a^4(\eta)} ~ \int^\infty_0  \sum_{s=\pm 1}\, \Big[\mathcal{V}^\dagger_s(-\vec{k},\eta)\,\vec{\alpha}\cdot \vec{k}\, \,\mathcal{V}_s(-\vec{k},\eta)  \Big]\, \frac{d^3k}{(2\pi)^3}\,,\label{Pvac} \\
P_{int} & = &   -\frac{1}{3 a^4(\eta)} ~ \int^\infty_0  \sum_{s=\pm 1}\,  \Big[A_{k,s}\,B^*_{k,s}  \,\mathcal{V}^\dagger_s(-\vec{k},\eta)\,\vec{\alpha}\cdot \vec{k} \,\mathcal{U}_s(\vec{k},\eta) +h.c. \Big]\, \frac{d^3k}{(2\pi)^3}\,,\label{Pint}\\
 P_{pp} & = &   \frac{1}{3  a^4(\eta)} ~ \int^\infty_0  \sum_{s=\pm 1}\, |B_{k,s}|^2 \Big[\mathcal{U}^\dagger_s(\vec{k},\eta)\,\vec{\alpha}\cdot \vec{k} \,\mathcal{U}_s(\vec{k},\eta)
 -  \mathcal{V}^\dagger_s(-\vec{k},\eta)\,\vec{\alpha}\cdot \vec{k} \,\mathcal{V}_s(-\vec{k},\eta)  \Big]\, \frac{d^3k}{(2\pi)^3}  \,,\nonumber \\ \label{Ppp1}    \eea where we have used the identities (\ref{bogorels},\ref{bogoconst}).

The terms  $\rho_{vac},P_{vac}$  are the  {vacuum} contributions during (RD); that this is the case should be clear from the fact that the spinors $\mathcal{V}$ are the solutions during (RD) with ``out'' boundary conditions associated with asymptotic anti-particle states.

The  terms  $\rho_{int},P_{int}$ describe the interference between positive and negative (asymptotic) energy solutions akin to the phenomenon of  \emph{Zitterbewegung}, and the last terms $\rho_{pp},P_{pp}$ describe the contributions from particle production with $|B_{k,s}|^2$ being identified as the  {distribution function} of the produced particles.

 \vspace{1mm}

 \textbf{Renormalization:}

 The expectation value of the energy momentum tensor in a gravitational background features ultraviolet divergences that must be renormalized.  The renormalization program  has been thoroughly studied in refs.\cite{bunch,pf,fh,hu,anderson,bir,mottola,birrell,parkerbook}, and extended for spin $1/2$ degrees of freedom in refs.\cite{rio,ferreiro,barbero,ghosh,landete}.

   The  {vacuum terms}, namely those for $B_{k,s} =0$, feature quartic, quadratic and logarithmic ultraviolet divergences that are renormalized by  {subtractions}. The   program to renormalize these divergences is well established and has been   implemented in refs.\cite{birrell,parkerbook,pf,fh,hu,bunch,anderson,mottola,bir,rio,ferreiro,barbero,ghosh,landete}. As discussed in detail in these references, the ultraviolet divergences are absorbed into renormalizations of the cosmological constant, Newton's constant $G$, and into the geometric tensors $H^{(1,2)}_{\mu \nu}$ which result from the variational derivative with respect to the metric of a gravitational action that includes higher curvature terms $\propto R^2, R^{\mu\nu} R_{\mu \nu}\cdots$. These higher curvature terms are added in the action multiplied by counterterms, which are then required to cancel the coefficients of the geometric tensors in such a way that the renormalized action is the Einstein-Hilbert action. The subtractions necessary to renormalize the (expectation value) of the energy momentum tensor, not only include the ultraviolet divergent terms but depending on the renormalization prescription may also subtract \emph{finite} terms. Therefore, {the finite renormalized energy momentum tensor is not unique} and depends on the renormalization prescription.

 Since $|B_{k,s}|$ is exponentially suppressed at large momentum, the interference and particle production contributions in the (RD) era are  { ultraviolet finite}, and originate, distinctly in the particle production mechanism,  whereas the  {vacuum} terms both during (RD) and inflation feature the ultraviolet  divergences and are independent of particle production as these are, simply, the zero point contributions. We renormalize the theory by \emph{completely subtracting the vacuum contribution to the energy momentum tensor, both during inflation and the (RD) stage}.

\subsection{Asymptotics: kinetic fluid form of $T^\mu_\nu$, (DM) abundance and equation of state.}

After subtraction of the vacuum terms, the renormalized $\langle 0|T^\mu_\nu|0\rangle$  {vanishes identically during inflation} and   features only the   particle production terms proportional to the Bogoliubov coefficient $B_k$ during (RD). In this latter era, the cosmological dynamics is dominated by the thermalized relativistic degrees of freedom of the standard model (and possibly beyond). Hence,  the contribution of the  (DM) degree of freedom can be neglected until it begins to dominate near   matter radiation equality. As discussed in section (\ref{subsec:adia}) the adiabatic approximation becomes reliable well before matter radiation equality for masses $m \gg 10^{-22}\,\mathrm{eV}$.

 In the adiabatic regime the exact solution of the  mode equation (\ref{heqn}) for the spinors $\mathcal{U},\mathcal{V}$ evolves into the WKB adiabatic solution (\ref{hout}) as discussed in section (\ref{subsec:RD}) and described in detail in appendix (\ref{app:ferad}).

Therefore, we can study the contribution of the energy momentum tensor in this regime by implementing the adiabatic expansion of the mode functions discussed in (\ref{subsec:adia}) and appendix (\ref{app:ferad}). We will show self-consistently below that the mass range of interest for (DM) abundance is certainly of the order of several $\mathrm{GeV}$ making the adiabatic approximation very reliable for $a(\eta) \gg 10^{-22}$  {well before} matter-radiation equality.

   The solution to  leading adiabatic order (zeroth order)   with ``out'' boundary conditions (\ref{heqn})  is given by (see Appendix (\ref{app:ferad}))
\be h_k(\eta) \propto e^{-i\int^\eta \omega_k(\eta')\,d\eta'} ~~;~~ \omega_k(\eta)= \sqrt{k^2+m^2H^2_R\eta^2}\,,  \label{hasi}\ee in terms of which, the zeroth order normalized spinor solutions are given by (we suppress the conformal time argument for ease of notation)
\be \mathcal{U}_s(\vec{k},\eta) = \frac{e^{-i\int^\eta \omega_k(\eta')\,d\eta'}}{\sqrt{2\omega_k(\omega_k+M)}} ~~\left( \begin{array}{c}
                                     (\omega_k+M )\, \xi_s\\
                                    k \, s \, \xi_s
                                  \end{array}\right)\,,  \label{Uspinrdasy}\ee

 \be \mathcal{V}_s(-\vec{k},\eta) =  \frac{e^{i\int^\eta \omega_k(\eta')\,d\eta'}}{\sqrt{2\omega_k(\omega_k+M)}}~~ \left( \begin{array}{c}
                                      -k \, s\,\xi_s\\
                                    (\omega_k + M)  \,   \xi_s
                                  \end{array}\right)\,.  \label{Vspinrdasy}\ee These spinors
are eigenstates of the instantaneous conformal time Dirac Hamiltonian $\vec{\alpha}\cdot {\vec{k}} + \gamma^0\,M(\eta)$ with eigenvalues $\pm \omega_k(\eta)$ respectively.

                                   The spinors $U^>,V^>$ are given in terms of these by the relations (\ref{Ugreat},\ref{Vgreat}), and the Bogoliubov coefficients
 are obtained in the previous section. For masses as large as a few $\mathrm{GeV}$ these solutions are an excellent approximation for $a(\eta) \gg 10^{-22} \gg a_{eq} \simeq 10^{-4}$, with corrections   $\ll \mathcal{O}(10^{-54})$ (see appendix (\ref{app:ferad})).

 In appendix (\ref{app:ferad}) we find up to second adiabatic order (see eqns. (\ref{rho2nd},\ref{pres2nd})),
 \be  V^\dagger(-\vec{k},\eta) \,\Sigma(\vec{k},\eta)\,V(-\vec{k},\eta) = -    U^\dagger(-\vec{k},\eta) \,\Sigma(\vec{k},\eta)\,U(-\vec{k},\eta) = - \omega\Bigg[1 -\frac{1}{8}\,\Big( \frac{a'}{ma^2}\Big)^2  \, \Big(\frac{k}{\gamma^2\,\omega} \Big)^2 \Bigg] \,, \label{rho2nd1}\ee

  \bea V^\dagger(-\vec{k},\eta) \,\vec{\alpha}\cdot \vec{k} \,V(-\vec{k},\eta) &  = &  -  U^\dagger(\vec{k},\eta) \,\vec{\alpha}\cdot \vec{k}\,U(\vec{k},\eta) = -\frac{k^2}{\omega_k}\Bigg\{1- \frac{1}{8\,\gamma^4}\,\Big( \frac{a'}{ma^2}\Big)^2 \, \nonumber \\ & \times & (1+\frac{1}{\gamma})\Big[1-\frac{1}{\gamma}+2(1+ \frac{1}{\gamma})\, \Big( \frac{\gamma-2}{(1+\gamma)^2}\Big)  \Big]
 \Bigg\} \label{pres2nd1}
 \eea

\noindent where $\gamma\equiv \sqrt{1+(k/ma)^2}$ is the local Lorentz factor.   The second terms inside the brackets are proportional to  $ \Big( \frac{a'}{ma^2}\Big)^2\simeq 10^{-54}/(m/\mathrm{eV})^2$ near matter radiation equality, and can be safely neglected. Therefore near matter radiation equality it is justified to  keep only the leading (zeroth) order term in the adiabatic expansion for the spinors.

 The leading adiabatic order spinors (\ref{Uspinrdasy},\ref{Vspinrdasy}) imply that the inteference terms feature the oscillatory factors
 \be e^{\pm 2i \int^\eta \omega_k(\eta')\,d\eta'} = e^{\pm 2i \int^t E_k(t')\,dt'}~~;~~ E_k(t) = \sqrt{ k^2_p(t)+m^2}~;~ k_p(t)= \frac{k}{a(\eta(t))}\,, \label{phases} \ee therefore, the interference terms average out by dephasing on (comoving) time scales $\lesssim 1/m$.

 Using the leading order spinors (\ref{Uspinrdasy},\ref{Vspinrdasy}) in the adiabatic regime during (RD) and the relations (\ref{bogorels},\ref{bogoconst}) among Bogoliubov coefficients, and,  neglecting the interference terms by averaging over their rapid oscillations \footnote{It turns out that at zeroth adiabatic order the interference terms $\rho_{int}$ vanish identically for each helicity separately, but not for $P_{int}$.}, we find

     \be \rho(\eta) =  \underbrace{-\frac{2}{2\pi^2 a^4(\eta)} ~ \int^\infty_0 k^2 dk\,\omega_k(\eta)
}_{zero~ point ~energy~density}+ \underbrace{\frac{4}{2\pi^2 a^4(\eta)} ~ \int^\infty_0 k^2 dk\,|B_{k,s}|^2~\omega_k(\eta)}_{particle~production}\,,\label{rhomatasy}  \ee

\be P(\eta) = \underbrace{-\frac{2}{6\pi^2 a^4(\eta)} ~ \int^\infty_0 k^2 dk\,\frac{k^2}{\omega_k(\eta)}
}_{zero~ point ~pressure}+ \underbrace{\frac{4}{6\pi^2 a^4(\eta)} ~ \int^\infty_0 k^2 dk\,|B_{k,s}|^2~\frac{k^2}{\omega_k(\eta)}}_{particle~production}\,.\label{Pmatasy}\ee The zero point energy density and pressure coincide with those obtained in \cite{rio}. It is straightforward to show covariant conservation of energy:
\be \dot{\rho}+ 3\,\frac{\dot{a}}{a}(P+\rho) = 0 \,, \label{covacons}\ee where the dot stands for derivative with respect to comoving time $t$. \emph{This identity holds separately for the vacuum and the particle production components}. We have purposely kept the vacuum terms to highlight the ultraviolet divergence. For example for the vacuum contribution to the energy density, expanding $\omega_k(\eta) \simeq k + M^2(\eta)/2k -3 M^4(\eta)/8k^3 +\cdots$ displays the quartic, quadratic and logarithmic divergences ubiquitous in the energy momentum tensor. Considering higher order adiabatic contributions to $\rho_{vac};P_{vac}$ it is found that these feature ultraviolet divergences up to quartic adiabatic order\cite{rio,ferreiro,ghosh,landete}, therefore the  {vacuum} contribution to the energy momentum tensor must be subtracted up to fourth adiabatic order.  However, because $|B_{k,s}|^2$ falls off exponentially at large momenta the contribution from particle production is  {ultraviolet finite}.

 We renormalize the energy momentum tensor by \emph{fully subtracting the zero point, vacuum  contributions to all orders in the adiabatic expansion}. See discussion on this point in section (\ref{sec:discussion}).  Because  covariant conservation (\ref{covacons}) holds separately for both the zero point and particle production contributions, the subtraction of the zero point contribution does not affect covariant conservation of the particle production term.

 Remarkably, the contribution from particle production is identified with the kinetic form of the energy density and pressure, where $|B_{k,s}|^2$ is the distribution function. The factors $4$ in the numerator of the contribution from particle production of (\ref{rhomatasy},\ref{Pmatasy}) arise from two polarizations and particle and antiparticle states, namely four degrees of freedom.

 The contributions from particle production are obtained by changing variables to $k = q \,\sqrt{mH_R}$. The Bogoliubov coefficient $|B_{k,s}|^2$ is solely a function of $q$ and is exponentially suppressed for $q > 1$; the product $q^2 |B_{k,s}|^2$ peaks at $q \simeq 0.6$ and drops-off exponentially. This behavior is displayed in figs. (\ref{fig:dist},\ref{fig:q2dist}). Writing
 \be \omega_k(\eta) = m\,a(\eta) \Big[ 1+ \frac{q^2}{a^2(\eta)}\,\frac{H_R}{m} \Big]^{1/2} \ee
  near matter radiation equality $a \simeq 10^{-4}$ and with $H_R/m \simeq 10^{-35}(\mathrm{eV})/m$ and $q \simeq 1$ it follows that we can safely approximate $\omega_k(\eta) \simeq m a(\eta)$ for $m  \geq 10^{-29}\,\mathrm{eV}$, implying that this is a non-relativistic species. Furthermore, consistently with the approximation of superhorizon modes,  the momentum integrals in (\ref{rhomatasy},\ref{Pmatasy}) must be cutoff at a scale $k_c \simeq 1/\eta_R = \sqrt{H_R H_{dS}}$, in terms of the variable $q$ this implies a cutoff $q_c \simeq \sqrt{H_{dS}/m} = 1/\varepsilon \gg 1$, however, because $|B_{k,s}|^2$ is exponentially suppressed for $q > 1$ the upper limit can be safely taken to infinity. Therefore, the particle production contributions to the energy density and pressure are given in terms of the ``emergent'' temperature $T_H$ by
  \be \rho_{pp}(\eta) = \frac{4\,\sqrt{2}\, m}{\sqrt{\pi}\, a^3(\eta)}\, \Big[m T_H\Big]^{3/2}\, \underbrace{\int^\infty_0 q^2\,|B(q)|^2\,dq }_{0.023} \,, \label{rhopp}\ee

  \be P_{pp}(\eta)= \frac{8\,\sqrt{2\pi}}{3 \,  m \,  a^5(\eta)}\, \Big[m T_H\Big]^{5/2}\,\underbrace{\int^\infty_0 q^4\,|B(q)|^2\,dq }_{0.01} \,. \label{Ppp} \ee

  If $|B(q)|^2$ in the integrands of (\ref{rhopp},\ref{Ppp}) is replaced by $ \frac{1}{4}\,e^{- \frac{k^2}{2mT_H}}$ we find the numerical values in these expressions to be $0.02\,,\,0.0095$ respectively. This   similarity has a remarkable consequence: noticing that the pre-factor $1/4$ that multiplies the  Maxwell-Boltzmann distribution in the asymptotic limit of $|B(q)|^2$ in (\ref{asybogoB})   cancels the factor $4$ in the particle-production contributions to (\ref{rhomatasy},\ref{Pmatasy}) leads us to conclude that the abundance and the equation of state differ only by $\simeq 10 \%$ from those obtained with a Maxwell-Boltzmann distribution function for a \emph{single  non-relativistic degree of freedom  at temperature $T_H$ and vanishing chemical potential} consistent with particles and antiparticles being produced with equal abundance.

Since the energy density redshifts as matter, we obtain (with $\rho_c = 3 H^2_0/8\pi G \simeq 0.4\,\times 10^{-10}\,(\mathrm{eV})^4$, and $\Omega_{dm}\simeq 0.25$)
\be \frac{\Omega_{pp}}{\Omega_{dm}} = a^3(\eta) \,\frac{\rho_{pp}(\eta) }{0.25\, \rho_c} \simeq   \Big( \frac{m}{ 3\,. 10^8\,\mathrm{GeV}}\Big)^{5/2} \,.\label{Oratio} \ee The equation of state parameter is given by
\be w(a) = \frac{P_{pp}(\eta)}{\rho_{pp}(\eta)} \simeq \frac{2\pi}{6 \, a^2(\eta)}\,\Big(\frac{T_H}{m} \Big)\,. \label{eosw}\ee These results differs by $\lesssim 10\%$ from those obtained for a single  a non-relativistic degree of freedom   with a Maxwell-Boltzmann distribution function  with a   temperature $T_H = H_R/2\pi$ since for non-relativistic particles $w \simeq <v^2>/3$ where $<v^2>$ is the velocity dispersion.

 These results suggest that this \emph{nearly thermal}  fermionic species with $m\simeq 10^8\,\mathrm{GeV}$ can be produced with    the  correct dark matter abundance and features the equation of state of  cold dark matter. Such value of the mass is consistent with our main approximation $m/H_{dS} \ll 1$ and the upper bound from Planck for $ 10^8\,\mathrm{GeV} \ll H_{dS} \lesssim  10^{13}\,\mathrm{GeV}$.

We also note the   following \emph{consistency aspects}:

\vspace{1mm}

\textbf{i:) range of momenta: } The integrals for the abundance and pressure are dominated by  the range $q \simeq 1$, namely, momenta $k \simeq \sqrt{mH_R}$. Therefore for momenta in this range it follows that  $k\eta_R \simeq \sqrt{m/H_{dS}} = \varepsilon  \ll 1$ for the values of $m$ that saturate the dark matter abundance and  $ H_{dS}$ in the above range. Therefore the integrals are dominated by wavelengths that are superhorizon at the end of inflation,  {consistently with one of our main approximations}.

\vspace{1mm}

\textbf{ii:) neglect of $\mathcal{O}(\varepsilon)$ terms: } In the calculation of the Bogoliubov coefficients we have neglected $\mathcal{O}(\varepsilon)$ terms both in the spinors and the functions $\mathcal{F},\mathcal{H}$. Since the integrals are dominated by the region $q\simeq 1 \gg \varepsilon$ and is suppressed at small momenta by phase space $\propto q^2$, neglecting these terms is warranted. Including these terms yields corrections of $\mathcal{O}(\varepsilon)$.

\section{Isocurvature perturbations:}\label{sec:iso}

\subsection{During inflation:}\label{subsed:infla}

 In the case of bosonic theories, adiabatic and entropy perturbations from inflation have been studied in refs.\cite{gordon,byrnes,bartolo} in the case where the bosonic fields associated with curvature and entropy perturbations both acquire   expectation values. Adiabatic and isocurvature perturbations result from linear combinations of fluctuations of the different bosonic fields around their respective expectation values. The case in which  {only} the inflaton field acquires an expectation value was addressed in ref.\cite{sena} within the context of (bosonic) superheavy dark matter produced during the inflationary era. This study  recognized that in the case in which the dark matter field does  {not} acquire an expectation value the treatment of isocurvature perturbations must be modified substantially. In particular, in absence of an expectation value for the entropy field there is no mixing between the fluctuations of this and the inflaton field   and no cross correlations between adiabatic and isocurvature perturbations to linear order. Several subtleties on the interpretation of isocurvature perturbations in the bosonic case have been discussed in ref.\cite{herring}.

 A similar situation arises in the case of fermionic fields since these cannot acquire an expectation value.  Fermionic isocurvature fluctuations were studied in ref.\cite{chungiso} within a model that couples fermions to another scalar field via a Yukawa coupling. Although the comoving isocurvature perturbations are defined by the following  correlation function of the (DM) energy momentum tensor
 \be \int \frac{d^3r}{(2\pi)^3} ~ e^{i\vec{k}\cdot\vec{r}}\, \langle \delta   (\vec{x}) ~\delta (\vec{x}+\vec{r}) \rangle \propto \mathcal{P}^{(dm)}(k)\,,   \label{Powdm}\ee where

  \bea \delta
  (\vec{x}) & = &  \frac{T^{(dm)}_{00}(\vec{x}) - \langle 0|T^{(dm)}_{00}(\vec{x})|0\rangle}{\rho^{(dm)}} \label{delt} \\ {\rho^{(dm)}} & = & \langle 0| T^{(dm)}_{00}(\vec{x})|0\rangle\,, \label{delrho}\eea the authors of ref.(\cite{chungiso}) only consider the correlations involving the composite operator $m \overline{\psi}\,\psi$ and take $\rho^{(dm)} \equiv m \langle \overline{\psi}\,\psi \rangle$ .

 The case that we study here departs from the model studied in (\cite{chungiso}) in several crucial aspects: \textbf{i:)} we do not consider  {any} coupling of the fermionic fields to any other bosonic field, \textbf{ii:)} the fermion field  in our case is in the Bunch-Davies vacuum state. As we discussed in detail in the previous section, we renormalize the energy momentum tensor by \emph{completely subtracting the vacuum contribution}, hence during inflation the \emph{renormalized} $\rho^{(dm)} =0$.  Therefore, the (DM) energy density perturbation (\ref{delrho}) cannot even be defined in the case that we study here.

 As discussed in section (\ref{sec:tmunu}) and in more detail in refs.\cite{birrell,pf,fh,hu,bunch,anderson,mottola,bir,rio,ferreiro,barbero,ghosh,landete} the expectation value of the energy momentum tensor features quartic, quadratic and logarithmic divergences, these are renormalized by {subtractions} absorbed in the counterterms in the gravitational action described in section (\ref{sec:tmunu}). The \emph{finite} part of $\langle 0|T_{\mu \nu}|0\rangle$ is not unique and depends on the renormalization prescription. When the field acquires an expectation value (background) the identification of $\rho^{(dm)}$ as that from the background energy momentum tensor, and $\delta $  as the contribution to  the energy momentum tensor \emph{linear in the fluctuations of the field}  are unambiguous. However, in absence of a background expectation value, the energy momentum tensor is at least quadratic in the fluctuations and its renormalization yields a finite part that depends on the renormalization procedure. In this scenario   $\rho^{(dm)}$ and  $\delta \rho$ are not uniquely defined. Because  we subtract the full expectation value of the energy momentum tensor in the Bunch-Davies vacuum state, it follows that $\rho^{(dm)}=0$ during inflation in our renormalization scheme.

 Therefore, neither the analysis of ref.(\cite{sena}) nor that of ref.(\cite{chungiso}) which specifically considers fermionic degrees of freedom, applies to the case that we study here.

 \subsection{Post inflation:}\label{subsed:postinfla}

  The discussion above has focused on the generation of entropy perturbations  {during inflation} and the applicability of the framework introduced in ref.\cite{sena,chungiso}. However, the relevant aspect is how  entropy (isocurvature) perturbations affect the temperature power spectrum of the (CMB). In the usual approach to cosmological perturbations, adiabatic and isocurvature perturbations during inflation provide the initial conditions of the respective perturbations upon horizon re-entry during the radiation (or matter) dominated era. As  discussed in detail in  refs.\cite{bartolo,byrnes}, the initial conditions of isorcurvature perturbations are determined by  the set of transfer functions discussed in ref.\cite{byrnes}. These, in turn, are proportional to the ``mixing'' (or correlation) angle which is determined by  the expectation value of the entropy field, and  \emph{vanishes identically} in the fermionic case.

 Furthermore, as we discussed above  the framework introduced in ref.\cite{chungiso} cannot be applied directly to the case that we study because the renormalization procedure that we follow subtracts the full expectation value of the energy momentum tensor in the Bunch-Davies vacuum during inflation. Therefore the background density vanishes identically in our case. This directly implies that the  initial conditions for isocurvature perturbations during the radiation dominated era  {cannot} be determined during the inflationary stage. In the radiation era the energy density and pressure feature  three contributions: the vacuum contribution is subtracted out in the renormalization procedure, the interference term is rapidly oscillating in the adiabatic regime ( for the energy density it vanishes at the leading adiabatic order) and therefore its expectation value averages out on short time scales, and   the contribution from particle production, which in the adiabatic regime features the kinetic fluid form. It is this latter term that is the relevant one (after renormalization) to understand dark matter perturbations, the distribution function is completely determined by the Bogoliubov coefficient $|B_k|^2$. The influence of isocurvature perturbations on the (CMB) is a result of solving the system of Einstein-Boltzmann equations for \emph{linear} cosmological   perturbations, in which  $|B_k|^2$ is the distribution function of the  {unperturbed} (DM) component, and   $\rho_{pp}$ (\ref{rhomatasy}) describes the \emph{background density}. This set of Einstein- Boltzmann equations  must be appended with initial conditions, which are   determined from the respective super-horizon perturbations at the end of inflation. From the above discussion, it is clear that in the case of fermions,  the proper initial conditions for isocurvature perturbations remain to be understood.


 \vspace{1mm}

 The corollary of this discussion is that a proper definition of the power spectrum of entropy perturbations in the case when the fields do  {not} acquire expectation values remains to be understood at a   deeper level. The caveats associated with the renormalization of the energy momentum tensor along with its correlations remain to be clarified in a consistent and unambiguous manner. These include a proper account of the fact that there is no natural manner to renormalize the divergences in a power spectrum obtained from the connected correlation function of the energy momentum tensor. These remain even when the zero point contribution to the energy density is completely subtracted. The contribution of zero point energy correlations to non-linear perturbations merits deeper scrutiny, \textit{since  even the fluctuations of the inflaton yield zero point contributions to the energy density and all other fields that are either produced or excited post-inflation   presumably  {also} contribute to the zero point energy density during inflation}.  A satisfactory resolution of these important issues, necessary   to   quantify reliably the impact of \emph{non-linear} entropy perturbations is still lacking,  and is clearly well beyond the scope of this study.

\section{Discussion:}\label{sec:discussion}

\textbf{On reheating:} The non-equilibrium reheating dynamics  leading  to a (RD) dominated era after inflation, is still a subject of much research.   Reheating dynamics is not universal, as a large body of studies show,  depending on  particular forms for the inflaton potential and the couplings of particles within and beyond the standard model to the inflaton and/or other degrees of freedom, thereby yielding model dependent descriptions with widely different time scales depending on generally unknown couplings and masses. See ref.\cite{reheat} for a review.

One of our main assumptions  is to focus on wavelengths that are superhorizon at the end of inflation. Two aspects of this assumption justify one of our main approximations, that of instantaneous reheating: the dynamics of the mode functions for these wave-vectors is on long time scales, hence insensitive to the reheating dynamics occurring on much shorter time scales. Furthermore, in principle, wavelengths larger than the particle horizon are causally disconnected from the   microphysical processes of thermalization. While this assumption  {seems} physically reasonable, it must be tested quantitatively. This requires studying a particular model of reheating dynamics, which however, would yield conclusions that would not be  universally valid. Perhaps a simple model that dynamically and  {continuously} interpolates (with continuous scale factor and Hubble rate) between a near de Sitter inflationary stage and a post-inflation (RD) stage would illuminate the validity of the instantaneous approximation. Such study would, undoubtedly,  require a substantial numerical effort to solve the mode equations during the transition and matching to the solutions in the subsequent (RD) era. Clearly such a study is beyond the scope of this article but merits further attention.

\vspace{1mm}

\textbf{Radiation density vs number of degrees of freedom.} During (RD) the Hubble rate is proportional to $\sqrt{g}$ with $g$ the effective number of ultrarelativistic degrees of freedom. In our analysis we took $\Omega_R$ to be the radiation component today, corresponding to $2$.  Therefore the value of $H_R$ (\ref{Hs}) and consequently of $T_H$ in (\ref{rhopp},\ref{rhopp}) scales as $\sqrt{g/2}$. For a fixed value of the mass $m$ the ratio (\ref{Oratio}) is multiplied by a factor $(g/2)^{3/4}$. If we assume only standard model degrees of freedom being thermalized after reheating, $g\simeq 100$,  in turn this implies that the numerator in  the bracket   of (\ref{Oratio}) is replaced by $m \rightarrow ~~\simeq   3\,m$. Hence the value of the mass that saturates the (DM) abundance is replaced by $m\simeq 10^8\,\mathrm{GeV}$, a simple rescaling by a factor $\mathcal{O}(1)$. Therefore, just taking the radiation fraction to be today's value yields a lower bound on the abundance and \emph{upper bound on the mass} that saturates the (DM) abundance. In any extension beyond the standard model $g$ will be larger, this implies that the ratio (\ref{Oratio}) must be multiplied by $ (\frac{g}{2})^{3/4}$ and the equation of state $w \rightarrow w \,\sqrt{g/2}$. The main conclusions remain the same with only a quantitative change: in the case of the standard model with $g\simeq 100$,  by a factor of $\mathcal{O}(1)$ in the mass bound and abundance, and a factor $\simeq 10$ in the equation of state, which, however will still describe cold dark matter.

\vspace{1mm}

\textbf{When are particles produced?:} This question does not have a unique answer. As discussed in refs.\cite{mottola,dunne} a \emph{time dependent} number operator for produced particles depends on the basis to define these particles. Different definitions or basis correspond to including higher adiabatic orders. While all these definitions yield the same number of particles \emph{asymptotically at long time} when the adiabatic approximation becomes reliable, they differ in the production dynamics during the non-adiabatic stages. This fact has been discussed in detail in ref.\cite{herring} and illustrated with various examples in ref.\cite{dunne}. We emphasize that we do  {not} introduce a number operator associated with a particular definition, instead we study the full energy momentum tensor and unambiguously extract the contribution of produced  {particles} asymptotically when the adiabatic approximation is very reliable. During the non-adiabatic stages at the end of inflation and early (RD) era, different definitions will yield very different dynamics.  Furthermore, as emphasized above, the continuity of the solutions of the Dirac equation, along with the continuity of the scale factor and Hubble rate at the transition from inflation to (RD) ensure that the energy momentum tensor is  {continuous} across the transition. Therefore there is no ``burst'' of particle production at the transition.

\textbf{Renormalization:}

 We have emphasized that the renormalization scheme that we implement subtracts completely the zero point contribution both during inflation and in (RD). Such scheme is, in fact, completely consistent with the usual working assumptions in (semiclassical) cosmology during and after inflation. For example, in the case of the inflaton, the background contribution to the energy momentum tensor is separated and assumed to drive inflation, the linearized perturbation of the energy momentum tensor around the background sources linear metric perturbations, but the quadratic and higher terms in the fluctuations are generically \emph{neglected}. However, these terms feature the ultraviolet divergences that must be renormalized; in not including this contribution in Einstein's equations, it is effectively completely subtracted out. Furthermore, if the standard model degrees of freedom are truly fundamental, they all contribute to the energy momentum tensor during inflation as well, when, presumably, all of these fields are in their (Bunch-Davies) vacuum state and their contribution amounts to zero point energy density and pressure. Not including their contribution in the dynamics of the geometry is tantamount to subtracting completely their contribution to the energy momentum tensor. In the (RD) era, the energy density of particles in thermal equilibrium is the expectation value of the Hamiltonian in the thermal density matrix and features the zero temperature zero point energy, which is ultraviolet divergent and is subtracted out. Therefore, the renormalization scheme that we adopt is consistent with the usual subtraction of zero point energies in cosmology.

\vspace{1mm}

\textbf{Comparison to previous work.} In refs. \cite{chungfer, kuzmin2} the gravitational production of fermions was studied. Our independent analysis agrees with some of these results while also presenting crucial differences. Regarding points of agreement: 1.) The long wavelength limit of the Bogoliubov coefficient $|B_{k,s}|^2 \rightarrow 1/2$ as $k\rightarrow 0$ is consistent with the Pauli blocking term in Fermi-Dirac distribution and confirms a similar limit in ref.\cite{chungfer}.  2.) For $m\ll H_{dS}$, we find that the final abundance does not feature any dependence on the inflationary stage in agreement with the results of ref.\cite{chungfer}. Moreover we find that this abundance has an overall dependence on the mass scale of the fermion species consistent with the result of \cite{kuzmin2}. 3.) The abundance we obtain saturates the necessary dark matter energy density at the same mass scale as obtained in \cite{chungfer, kuzmin2}. We note that this is likely a consequence of $m$ and $H_R$ (the Hubble scale during RD) being the only relevant scales in the $m\ll H_{dS}$ scenario.

However, we also recognize important differences in our results from the literature: 1.) For large $k$, our distribution given by (\ref{almMB}) is in stark contrast to the behavior shown in fig. (1) of \cite{chungfer} (see $m \ll H_{dS}$ case). The authors quote a power law $|B_{k,s}|^2\simeq k^{-4}$ for   large $k$ regarding modes which are super-horizon at the end of inflation. This disagrees with our description of Maxwell-Boltzmann-like exponential behavior for these same modes. We do not understand the origin of this important difference. 2.) The matching conditions employed in \cite{chungfer} for enforcing continuity of the mode functions from inflation to RD are quite distinct from our procedure. We perform an in-out calculation, analytically solving the equations of motion for the mode functions in the $m \ll H_{dS}$ limit. While the solution in the inflationary regime is fixed by our initial condition (BD vacuum, in-state), the solution in RD regime must be the  {general solution} constructed out of a linearly-independent combination of solutions. Each  {particular solution} of this linear combination smoothly, asymptotically matches onto our adiabatic, out-states (i.e. our particle and anti-particle states with associated u-type and v-type spinors.) Thus our matching condition leads to  (\ref{Ugreat},\ref{Vgreat}) where a u-type spinor in inflation is a linear combination of u-type and v-type spinors in RD. In \cite{chungfer}, the u-type spinors of inflation are matched with solely u-type spinors during RD. We believe this important difference is a result of the authors in \cite{chungfer} using a time-dependent particle number which we  {do not} resort to for reasons discussed above. 3.) Neither \cite{chungfer} nor \cite{kuzmin2} obtain the energy density or pressure. Given their result of $|B|^2 \propto 1/k^4$ these quantities would depend logarithmically on $H_{dS}$ as discussed above. 4.) In \cite{chungfer}, the authors introduce an upper bound on the mass of the dark fermion candidate for their calculation to be consistent. When this upper bound is imposed the produced particle abundance becomes negligible and cannot saturate the necessary dark matter abundance. Conversely, in our calculation we do not find any upper bound more restrictive than $m \ll H_{dS}$ making our result (\ref{Oratio}) robust and general. To be clear, we have shown self-consistently, that only the low momentum, superhorizon modes contribute to the abundance and equation of state; therefore, our instantaneous reheating approximation is well-justified and our results are insensitive to any reheating model-dependent parameters.

\vspace{1mm}

\textbf{Fermions vs Bosons:}
Ref. \cite{herring} studied the cosmological particle production of scalar particles both minimally (MC) and conformally (CC) coupled to gravity focusing on ultralight dark matter candidates ($m \ll H_{dS}$) under the same assumption of an instantaneous transition to the (RD) era. Thus we can now compare the results of that study with the analysis conducted here. The result for the abundance in the bosonic (CC)   case given by eqn. (V.42) in ref.\cite{herring} is remarkably similar to the fermionic abundance (\ref{Oratio}) after the proper rescaling of the energy units. However, the similarity of the results conceals very important differences between the bosonic and fermionic cases.

\begin{figure}[ht!]
\begin{center}
	\includegraphics[height=4in, width=4in,keepaspectratio=true]{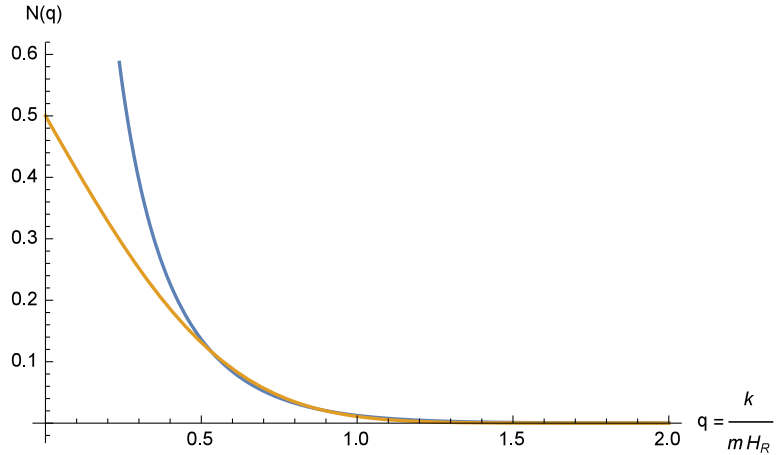}
	\caption{The produced particle distributions  N(q) where $q = \frac{k}{m H_R}$ for conformally coupled bosons (blue) and fermions (gold). (Colors available online.) }
	\label{fig:ccvsfermi}
\end{center}
\end{figure}

 First we highlight the differences in the produced particle distribution (focusing on the asymptotic regimes) using equations (\ref{asybogoB}) and (III.63, III.69 from \cite{herring}):
\begin{align*}
k \ll& \sqrt{m H_R}:  \\
N_{k,s} \simeq \frac{1}{2} \;\; \text{(Fermions)} \;\;\;
N_{k} \propto &\frac{1}{k} \;\; \text{(CC)} \;\;\;
N_{k} \propto \frac{1}{k^3} \;\; \text{(MC)}
\end{align*}
\begin{align*}
k \gg& \sqrt{m H_R}: \\
N_{k,s} \simeq e^{-\frac{ k^2}{2 m T_H}} \;\; \text{(Fermions)} \;\;\;
N_{k}  \propto &\frac{1}{k^8} \;\; \text{(CC)} \;\;\;
N_{k} \propto \frac{1}{k^4} \;\; \text{(MC)}
\end{align*}
In all three cases the distribution function is peaked at low co-moving wave vectors. However, in the fermion case it is bound by Pauli blocking at very low momentum reaching the maximum value $1/2$ and  is \emph{exponentially} suppressed for large momenta by the thermal factor, while in the bosonic case the distribution function diverges as a power law in the infrared  and falls off with  a different power law at large momentum.  The distribution functions for the  bosonic (CC) and the fermionic case are displayed in Fig.   (\ref{fig:ccvsfermi})).\par

For minimally coupled bosons extracting the matter-like contribution to the energy density (the component which redshifts as $1/a^3(\eta)$) requires introducing an upper integration bound ($k \lesssim m a_{eq}$). When combined with the stipulation of superhorizon modes at the end of inflation ($k\eta_{R} \ll 1$), this results in an upper mass bound $m \lesssim \frac{1}{a_{eq}\eta_R}\simeq 0.01 \,\mathrm{eV}$ for ultralight, non-adiabatic particle production. One  \emph{does not obtain} this bound in the case of fermions because the exponential suppression of the distribution function permits one to integrate over all momenta self-consistently as discussed above. Thus, one can consider non-adiabatic particle production of TeV-scale fermions (or even higher) with only $m \ll H_{dS}$ required. \par

The comparison between fermions and conformally coupled bosons is more apt as both cases obey the same de Sitter mode equation for superhorizon modes (for $m \ll H_{dS}$) (\ref{desieqn}). This is unsurprising since in either case there is no direct coupling between the scalar curvature and the quantum fields. However, despite these similarities with conformally coupled bosons, the fermion mode functions  {do not} feature an infrared enhancement (compare (\ref{fetaR}) with (III.12) of \cite{herring}) unlike those of either bosonic case. This discrepancy is   a consequence of the normalization of the fermion spinors, and ultimately the canonical anticommutation relations and Pauli blocking,  and explains the differences in the low momentum behavior of the distribution functions (see Fig. (\ref{fig:ccvsfermi})). These differences in the distribution function are not a just a formal issue, the  {moments} of the distribution function will be very different. These moments enter in the Boltzmann hierarchy for the coupled radiation-matter-gravitational perturbations during the (RD) era prior to matter radiation equality. Therefore, despite the similarity in the abundance between the bosonic (CC) and fermionic cases, we expect substantial modifications in the transfer functions obtained from Boltzmann codes from the two very different distribution functions. These aspects remain to be studied further, however they are beyond the scope of this article.

\vspace{1mm}

\vspace{1mm}

\vspace{1mm}

\textbf{On thermality:} A noteworthy aspect of our result is a near thermal distribution of produced particles with the distribution function (\ref{almMB}). It yields a near thermal abundance and equation of state very similar to that of one non-relativitistic degree of freedom with a Maxwell-Boltzmann distribution at the ``emergent temperature'' $T_H$. The surprising emergence of this temperature is unexpected because an (RD) cosmology does not feature an event horizon, hence this temperature cannot be identified with   Gibbons-Hawking radiation\cite{gib}. Aspects of thermality in the distribution of particles produced via cosmological expansion were also revealed in early work in refs.\cite{audretsch,hartle} in very different cosmological settings. Ref.\cite{audretsch} studied cosmological production in a radiation dominated cosmology \emph{including and extending the singularity}, the authors find  a distribution of particles that is also non-relativistic   with en effective temperature (see also \cite{birrell} section 3.5).  A similar conclusion was reached in ref.\cite{hartle} that studied cosmological particle production in a path-integral framework but with a different cosmological model with scale factor $\propto t$ (t being comoving time), and an effective temperature that is  $\propto a(t)$. Our results apply to a very different situation since we consider fermion fields, initial ``in'' conditions during inflation, and match onto (RD). Furthermore, the ``emergent temperature'' that we find, $T_H$,  is independent of time. Therefore, while the results of the early references \cite{audretsch,hartle} suggest a general ``thermality'' aspect of the distribution of produced particles, the relationship to our results, if any, and the physical origin of the near thermal spectrum is not clear to us.

\vspace{1mm}

\textbf{Pair annihilation into gravitons:} Because we are considering that the dark matter candidate only couples to gravity, the process of particle anti-particle pairs   annihilating into gravitons   could lead to a depletion of  the (DM) abundance. Fundamentally, to understand the dynamics of this process one would set up a Boltzmann-like equation with a loss term determined by the annihilation process, assuming a negligible abundance of gravitons one can, in principle, neglect the inverse process. Such an equation would feature the six-dimensional momentum integrals of the distribution functions for the annihilating pairs multiplied by the transition probability obtained from the time evolution via the interaction picture. For the case of annihilation into gravitons,  writing the metric as $g_{\mu \nu} = \overline{g}_{\mu \nu} - \frac{h_{\mu \nu}}{M_{Pl}}$ with $\overline{g}_{\mu \nu}$ being the background metric and $h_{\mu \nu}$ the canonically normalized quantum fluctuations of the gravitational field, the coupling to gravitons is given by 
\be  \frac{h_{\mu \nu}}{M_{Pl}}\,T^{\mu \nu} \,. \label{gravcoup}\ee  This is the interaction vertex that is required in the interaction picture to obtain the transition probability. The usual implementation of S-matrix theory taking the long time limit to obtain the transition probability (per unit time) is unreliable for the following conceptual and technical  reasons.

 In transverse-traceless gauge, the graviton field is expanded into canonical creation and annihilation operators for each polarization ($+, \times$) and mode functions solutions of the Klein-Gordon equation for a massless field minimally coupled to gravity.    Conceptually  the notion of particles is physically unambiguous only in the asymptotic long time regime. Using the basis of fermion  ``out'' states entails that during the \emph{non-adiabatic} regime, during which most of the particle production takes place, the mode functions for these states are the parabolic cylinder functions,  not the usual Minkowski type exponentials $e^{\pm i\omega t}$. The mode functions for gravitons during (RD) are actually of the form $e^{\pm i k\eta}/a(\eta)$, and while spatial momentum is conserved (in a spatially flat metric with three space-like Killing vectors) energy is \emph{not} conserved. Taking the infinite time limit in the transition amplitudes as is implicit in an S-matrix calculation, is obviously unreliable in a rapidly expanding cosmological setting. Therefore a  calculation even to lowest order is very different from that in Minkowski space time and confronts daunting technical challenges, that to the best or our knowledge have not yet been discussed, much less worked out in the literature. For example, even during the adiabatic regime the correct assessment of particle decay in an expanding cosmology is technically challenging as can be gleaned from the work in refs. \cite{decay1,decay2} with results which are generally very different from those expected in Minkowski space time. A similar calculation for annihilation has not been carried out in the literature even in the adiabatic regime.

 Therefore, in light of these conceptual and technical challenges it should be clear that a reliable quantitative assessment of the influence of pair annihilation in the expanding cosmology is well outside the main scope of this article and must await the development of new techniques. Here we can at best    provide a  very preliminary and rough estimate for the depletion from pair annihilation into two gravitons at second order in the interaction based on a Minkowski intuition and the main scales involved.  Assuming (without warrant) that Minkowski-like dynamics provides a useful guide,  the main ingredients in this argument are the following: i)  there are initially no gravitons so that the inverse process does not occur (this by itself is a major assumption since gravitons are produced during and post inflation), ii) the strength of the vertex (for the spatial components) (\ref{gravcoup}) is determined by the typical momentum in the distribution functions $\simeq \sqrt{mH_R}$,  and the typical energy scale   $\simeq m$,  which we take as the main scale for the fermionic degrees of freedom. Therefore the effective coupling in this vertex is $\simeq m/M_{pl} \simeq 10^{-10}$ for $m\simeq 10^8 \,\mathrm{GeV}$ which saturates the bound for (DM) abundance (\ref{Oratio}). The intermediate state yields  a fermion propagator with a typical scale $1/m$,  the comoving number density is $n \simeq (mH_R)^{3/2}$ and the probability for pair annihilation is $\propto n^2$. Thus  we are led to conclude that on dimensional grounds, and within the assumption of the validity of  a Minkowski space-time estimate,  at second order in the vertex
 \be \frac{\delta n}{\delta t} \simeq \frac{n^2}{m^2}\, \Big(\frac{m}{M_{Pl}} \Big)^4\,,  \ee integrating this expression during a time interval $\Delta t \simeq 1/H_R$ yielding a total depletion $\Delta n =   \delta n/H_R \delta t$, we find
 \be \frac{\Delta n}{n} \simeq \Big( \frac{H_R}{m}\Big)^{1/2} \,\Big(\frac{m}{M_{Pl}} \Big)^4 \simeq 10^{-65} \label{deple} \,.\ee Therefore,   this assessment suggests that pair annihilation into gravitons will not affect the abundance. However this result should be interpreted as a guide with all of the caveats discussed above, a more detailed assessment with new methods  that can describe consistently the time evolution of annihilation during the non-adiabatic regime is needed. Such method should not rely on the usual S-matrix approximations of taking the long time limit with manifest energy conservation and should input the correct mode functions. The program to develop these new methods and applying them consistently to the calculation of pair annihilation  is well beyond the scope of this article and merits a detailed study.

\section{Conclusions and further questions:}\label{sec:conclusions}
We have studied the gravitational production of fermionic dark matter during inflation and radiation domination under a minimal set of assumptions: i) its mass $m$ is much smaller than the Hubble scale during inflation, described as de Sitter space time, ii) only interact gravitationally, iii) fermions are in the Bunch-Davies vacuum state during inflation, iv) focus on wavelengths that are well outside the Hubble radius at the end of inflation, v) a rapid transition from inflation to (RD).

 We solve exactly the Dirac equation during inflation and radiation domination with ``in'' and ``out'' boundary conditions and match the solutions at the transition.   Particle-antiparticle pairs  are produced non-adiabatically with a distribution function $|B(k)|^2 = \frac{1}{2}\Big[1-\big(1-e^{-\frac{k^2}{2mT_H}}\big)^{1/2} \Big] $ exhibiting an \emph{emergent temperature} $T_H= H_0 \sqrt{\Omega_R}/2\pi \simeq 10^{-36}\,\mathrm{eV}$ with $H_0,\Omega_R$ the Hubble expansion rate and radiation fraction today respectively. This distribution function is remarkably similar to a Maxwell-Boltzmann distribution for a non-relativistic species with vanishing chemical potential, in agreement with the fact that particles and antiparticles are produced with the same distribution.

With the exact solution we obtain the full energy momentum tensor, discuss in detail its renormalization and extract unambiguously the contribution from particle production near matter radiation equality. We show that after renormalization this contribution features the kinetic-fluid form with $|B(k)|^2$ as the distribution function. We obtain the energy density $\rho_{pp}$, pressure $P_{pp}$  and equation of state parameter $w(a)$ of the produced particles   factor $a$
\bea \rho_{pp} & = & 0.074\,\frac{m}{a^3}\,\big[m\,T_H \big]^{3/2} \nonumber \\
P_{pp}& = & 0.067\,\frac{\big[m\,T_H \big]^{5/2}}{m\,a^5} \nonumber \\
w(a) & = & \frac{P_{pp}}{\rho_{pp}} \simeq \Big[ \frac{T_H}{m\,a^2}\Big]  \,, \nonumber \eea where $a$ is the scale factor. Remarkably these correspond to a \emph{nearly thermal} non-relativistic species in equilibrium at temperature $T_H$ and vanishing chemical potential,  with the equation of state function related to the velocity dispersion for this species as $w(a) \simeq \langle V^2 \rangle/3$. The departure from an  {exactly} thermal non-relativistic  single species with a Maxwell-Boltzmann distribution at temperature $T_H$ is $\lesssim 10\%$. The reason behind this small discrepancy is the behavior of $|B(k)|^2$ as $k \rightarrow 0$.

The ratio of  the abundance of produced particles to the dark matter abundance  is given by
\be \frac{\Omega_{pp}}{\Omega_{dm}} =  \Big( \frac{m}{ 3\,. 10^8\,\mathrm{GeV}}\Big)^{5/2} \,.  \ee Therefore, a fermionic particle with mass $\simeq 10^8 \,\mathrm{GeV}$ can be produced gravitationally, with the correct dark matter abundance and constitutes cold dark matter.

The integrals yielding $\rho_{pp},P_{pp}$ are dominated by wavevectors $k \lesssim  \sqrt{2mT_H}$ and these correspond to wavelengths that are well outside the Hubble radius at the end of inflation, confirming self-consistently the validity   of the main approximation of focusing solely on these wavevectors for the matching conditions from inflation to (RD),

We discuss important aspects of renormalization during both the inflationary and (RD) era which imply subtle uncertainties associated with an unambiguous determination of isocurvature perturbations from gravitationally produced fermions. These uncertainties are not only a characteristic of fermionic degrees of freedom, but apply generally to fields that do  {not} acquire an expectation value during inflation and whose energy momentum tensor feature ultraviolet divergences of the zero point contributions that must be renormalized by proper subtractions. These subtractions depend on the particular renormalization scheme, therefore  the finite part of the energy density and pressure arising from the renormalization of the zero point contributions would be scheme dependent. Our procedure is to subtract completely the zero point contributions, and we argue that this procedure is implicitly implemented in all treatments of inflationary dynamics. However, the corollary of this subtraction is that the  initial conditions for isocurvature perturbations during radiation or matter domination \emph{cannot} be defined during the inflationary epoch. The resolution of these aspects  remains a subject of further study.

   The origin of thermality in the distribution function is also an aspect that merits further understanding, since it cannot be identified with a Gibbons-Hawking temperature because an (RD) cosmology does not feature an event horizon.

\vspace{1mm}

\appendix

\section{Majorana fermions.}\label{app:majorana}

In this appendix we gather the main features for the quantization of Majorana fermions. With the solutions of the Dirac equation obtained in section (\ref{sec:model}), we construct self-conjugate Majorana fermions as follows.

Introducing
\be w(k,\eta) =  i \frac{f'_k(\eta)}{f_k(\eta)}+M(\eta) \label{wofketa}\ee where $' = d/d \eta$, the Dirac spinors are written as
\be U_\lambda(\vk,\eta) = N  \,f_k(\eta)\, \left(
                            \begin{array}{c}
                              w(k,\eta)\, \chi_\lambda \\
                              \vec{\sigma}\cdot \vec{k}\, \chi_\lambda \\
                            \end{array}
                          \right) ~~;~~ \chi_1 = \left(
                                                   \begin{array}{c}
                                                     1 \\
                                                     0 \\
                                                   \end{array}
                                                 \right) \;; \; \chi_2 = \left(
                                                                       \begin{array}{c}
                                                                         0 \\
                                                                         1 \\
                                                                       \end{array}
                                                                     \right) \,,
 \label{Uspinorsolm} \ee  and

 \be V_\lambda(\vk,\eta) = N  \,f^*_k(\eta)\, \left(
                            \begin{array}{c}
                               \vec{\sigma}\cdot \vec{k}\, \varphi_\lambda \\
                               w^*(k,\eta)\, \varphi_\lambda  \\
                            \end{array}
                          \right) ~~;~~ \varphi_1 = \left(
                                                   \begin{array}{c}
                                                    0 \\
                                                     1 \\
                                                   \end{array}
                                                 \right) \;; \; \varphi_2 = -\left(
                                                                       \begin{array}{c}
                                                                         1 \\
                                                                         0 \\
                                                                       \end{array}
                                                                     \right) \,,
 \label{Vspinorsolm} \ee  where $\lambda = 1,2$.  These spinors are normalized
 \be U^\dagger_\lambda U_{\lambda'} = V^\dagger_\lambda V_{\lambda'} = \delta_{\lambda,\lambda'}\,, \ee   yielding the same normalization factor as obtained in section (\ref{sec:model}), and fulfill the orthogonality condition
 \be U^\dagger_\lambda(\vec{k},\eta) V_{\lambda'}(-\vec{k},\eta)  = 0 ~~;~~ \lambda,\lambda' = 1,2\,. \ee

    It is straightforward to confirm that  the $U$ and $V$ spinors (\ref{Uspinorsolm}, \ref{Vspinorsolm}) obey the charge conjugation relation
 \be i\gamma^2 U^*_\lambda(\vk,\eta) = V_\lambda(\vk,\eta) ~~:~~ i\gamma^2 V^*_\lambda(\vk,\eta) = U_\lambda(\vk,\eta) ~~;~~ \lambda = 1,2 \,. \label{chargeconj}\ee In terms of these spinor solutions we can construct Majorana (charge self-conjugate) fields obeying\footnote{We set the Majorana phase to zero as it is not relevant for the discussion.}
 \be \psi^c_M(\vx,\eta) = C (\overline{\psi}_M(\vx,\eta))^T  = \psi_M(\vx,\eta) ~~;~~ C = i\gamma^2\gamma^0 \label{majorana} \ee and given by
\be
\psi_M(\vec{x},\eta) =    \frac{1}{\sqrt{V}}
\sum_{\vec{k},\lambda}\,   \left[b_{\vec{k},\lambda}\, U_{\lambda}(\vec{k},\eta)\,e^{i \vec{k}\cdot
\vec{x}}+
b^{\dagger}_{\vec{k},\lambda}\, V_{\lambda}(\vec{k},\eta)\,e^{-i \vec{k}\cdot
\vec{x}}\right] \; .
\label{psiexmajo}
\ee
In the case of Majorana fields the     Lagrangian density, and the energy momentum tensor  must be multiplied by a factor $1/2$ since a Majorana field has half the number of degrees of freedom of the Dirac field. Furthermore, one can take linear combinations of the Weyl spinors $\chi_{1,2};\varphi_{1,2}$ and construct helicity eigenstates. The steps leading to the final form of the abundance and equation of state are the same as for the Dirac case, with the only difference being a factor $2$ instead of the factor $4$ because for a Majorana field particles are the same as antiparticles, thereby halving the number of degrees of freedom.

\section{Properties of the solution (\ref{soluh}):}\label{app:solutionRD}

During the (RD) stage the solution of the mode functions is given by (\ref{soluh}) with $z,\alpha$ given by eqns. (\ref{dimcombos},\ref{solus}) respectively. Using the properties of the parabolic cylinder functions\cite{as,nist,bateman,magnus} we find for $|z| \gg |\alpha|$
\be D_\alpha[\sqrt{2}e^{i\pi/4}\,z] \simeq e^{-i\frac{z^2}{2}}\,\Big[\sqrt{2}e^{i\pi/4}\,z  \Big]^{\alpha}\Big[1+\cdots\Big]\,, \label{largez}\ee
and for $|\alpha| \gg |z| \gg 1$
\be D_\alpha[\sqrt{2}e^{i\pi / 4}\,z] \simeq \frac{e^{\pi q^2/8}}{\sqrt{2q^2}}\,e^{-izq}\,\Big[1+ \cdots\Big] \,, \label{largealfa}\ee
Up to an overall phase and normalization, these limits describe the asymptotic WKB solution (\ref{hasi}) yielding the spinor solutions (\ref{Uspinrdasy},\ref{Vspinrdasy}) valid in the adiabatic limit. Note that the term $e^{\pi q^2/8}$ cancels the normalization factor (\ref{normar})  in this limit, yielding the correct normalization of the spinors (\ref{Uspinrdasy},\ref{Vspinrdasy}).

Since the matching condition is evaluated at $\eta_R$, in terms of the variable $z$ it follows that $z_R = \sqrt{m H_R}/\sqrt{H_R H_{dS}} = \varepsilon \ll 1$.  For $z\simeq 0$ we find
\be D_\alpha[\vartheta] = D_\alpha[0]\,\Big[1 + \frac{1}{2}(1+iq^2)\vartheta^2 +\cdots \Big]+ \frac{d}{d\vartheta}D_\alpha[\vartheta]\Big|_{\vartheta=0}\,\vartheta\, \Big[1 + \frac{1}{12}(1+iq^2)\vartheta^2 +\cdots \Big]\,. \label{smallz}\ee
 At the transition time $\eta = \eta_R$ it follows that
\be  q^2 z^2_R =\frac{k^2}{m H_R} \, m H_R\,\eta^2_R = (k\eta_R)^2 \ll 1 \,, \label{atetaR}   \ee
 for superhorizon wavelengths at the end of inflation. Therefore, for $\varepsilon \ll 1$, and $ k\eta_R \ll 1$ we can reliably  approximate
 \be D_{\alpha}(z_R) \simeq D_{\alpha}(0) ~~;~~ \frac{d}{d\eta}D_{\alpha}(z)|_{z_R} \simeq \frac{d}{d\eta}D_{\alpha}(z)|_{z=0}\,. \label{zetazero}   \ee

 \section{Calculation of $|B_{k,s}|^2$}\label{app:BogoBk}
 Neglecting terms of $\mathcal{O}(\varepsilon)$ the spinors $\mathcal{U},\mathcal{V}$ are given by
 \be \mathcal{U}_s(\vec{k},\eta) \simeq \widetilde{N}\,\left( \begin{array}{c}
                                      ih'_k \, \xi_s\\
                                    k\,h_k \, s \, \xi_s
                                  \end{array}\right)\,,  \label{Uspinrd2}\ee

 \be \mathcal{V}_s(-\vec{k},\eta) \simeq \widetilde{N}\,\left( \begin{array}{c}
                                      -k\,h^*_k \, s\,\xi_s\\
                                     -ih^{*'}_k    \,   \xi_s
                                  \end{array}\right)\,.  \label{Vspinrd2}\ee The normalization constant is determined by
\be |\widetilde{N}|^2 \Big[|h'_k(\eta_R|^2 +k^2 |h_k(\eta_R|^2\Big] = 1 \,, \ee       using the
identities\cite{as}
\be \Big|\Gamma(\frac{1}{2}+i \frac{q^2}{4})\Big|^2 = \frac{\pi}{\cosh[\pi q^2/4]} ~~; ~~ \Big|\Gamma(1+i \frac{q^2}{4})\Big|^2 = \frac{\pi (q^2/4)}{\sinh[\pi q^2/4)]} \,.\label{ids}\ee we find
\be |\widetilde{N}|^2 = \frac{e^{-\pi q^2/4}}{2mH_R}\,. \label{Nwidet}\ee

The final form of $|B_{k,s}|^2$ is given by eqn. (\ref{bogofin1}), neglecting terms of $\mathcal{O}(\varepsilon)$ as discussed in section (\ref{subsec:bogos}) we find

\be \mathcal{H}^*_k(\eta_R)\,h_k(\eta_R)+\mathcal{H}_k(\eta_R)\,h^*_k(\eta_R) = \frac{e^{i\pi/4}}{\Gamma(\frac{1}{2}-i\frac{q^2}{4})\,\Gamma(1+i\frac{q^2}{4})}+ \frac{e^{-i\pi/4}}{\Gamma(\frac{1}{2}+i\frac{q^2}{4})\,\Gamma(1-i\frac{q^2}{4})}\,. \ee
Using the doubling formula for Gamma functions\cite{as,bateman,magnus}, and defining $q^2/4 \equiv y$
\be \Gamma(1+iy)\,\Gamma(\frac{1}{2}-iy) = 2\sqrt{\pi}\,y\, e^{i\pi/2}\,e^{2iy\, ln(2)}\, \Gamma(-2iy) \frac{\Gamma(iy)}{\Gamma(-iy)}\,, \label{formu1}\ee and writing
\be \Gamma(-2iy) \frac{\Gamma(iy)}{\Gamma(-iy)} = \Big|\Gamma(-2iy) \Big| \,e^{i\Phi(y)} ~~;~~ \Big|\Gamma(-2iy) \Big| = \Bigg[\frac{\pi}{2y\,\sinh[2\pi y]} \Bigg]^{1/2}\,,  \ee  where
\be \Phi(y) = \mathrm{Im} \Bigg\{\ln[\Gamma(-2iy)] + \ln[\Gamma(iy)]-\ln[\Gamma(-iy)]  \Bigg\} \,. \label{phian}\ee
Using the identity\cite{as,bateman,magnus}
\be \ln[\Gamma(z)] = \big(z-\frac{1}{2} \big)\,\ln(z) -z+ \frac{1}{2} \ln(2\pi) + \int^\infty_0 \Bigg[ \frac{1}{2} -\frac{1}{t} + \frac{1}{e^t -1}\Bigg]\,\frac{e^{-iz}}{t}\,dt \,,\label{form2}  \ee we find
\be \Phi(y) = -\frac{\pi}{4}-2y\ln(2)\,,  \label{fifini}\ee which, combined with the normalization factor (\ref{normar}) yields
  the final result (\ref{BogoBk}).

  \section{Adiabatic expansion for Fermi fields:}\label{app:ferad}
  In this appendix we provide a discussion of the adiabatic approach to fermionic degrees of freedom which is  alternative to the framework discussed in refs.\cite{rio,ferreiro,barbero,ghosh,landete}, with the advantage that it yields more compact expressions for the energy density and pressure. We write generically the spinors as $U$, $V$ with the understanding that during (RD) these are to be identified with the solutions  $\mathcal{U}\,;\,\mathcal{V}$.

 Consider the mode equation  (we suppress the momentum label and conformal time arguments for ease of
 notation)
 \be h^{''}+ (\omega^2-iM')h =0 \label{modeh}\ee and propose the solution
 \be h(\eta) = e^{-i\int^\eta \Omega(\eta')\,d\eta'} ~~;~~ \Omega= \Omega_R + i\Omega_I \,. \label{solwkb}\ee Introducing this ansatz into the mode equation (\ref{modeh}) yields
 \be \Omega^2 + i\Omega' - \omega^2 + iM' =0 \,, \label{eqnOme}\ee separating the real and imaginary parts yields the coupled system of equations
 \bea && \Omega^2_R - \Omega^2_I - \Omega^{'}_I -\omega^2 = 0 \,\label{realOm}\\
 && 2\Omega_R\Omega_I + (\Omega^{'}_R+M')= 0 ~~ \Rightarrow ~~ \Omega_I = -\frac{(\Omega^{'}_R+M')}{2\Omega_R}\,.  \label{ImagOm}\eea The above equations can be solved in a consistent adiabatic expansion in derivatives of $\omega, M$ with respect to conformal time.
  A corollary of these equations is that $\Omega_R, \Omega_I$ feature  an adiabatic expansion {even} and  {odd}   in the number of derivatives (adiabatic order) respectively, with
  \be \Omega^{(0)}_R = \omega~;~ \Omega^{(0)}_I = 0 ~~;~~  \Omega^{(1)}_R=0 ~;~ \Omega^{(1)}_I = -\frac{(\omega'+M')}{2\omega}~~;~~ \Omega^{(2)}_R = \frac{(\Omega^{(1)}_I)^2 + (\Omega^{(1)}_I)'}{2\omega}~;~\Omega^{(2)}_I = 0 \cdots \,.  \label{lowords}\ee To highlight the nature of the adiabatic expansion, consider the dimensionless ratio

   \be \frac{\Omega^{(1)}_I}{\Omega^{(0)}_R} =  - \frac{a'}{m\,a^2}\Big[\frac{1}{\gamma^2}+ \frac{1}{\gamma^3} \Big]\,\label{adi}\ee
  \be \gamma = \sqrt{\frac{k^2}{m^2 a^2}+1} \geq 1  \label{galor}\ee is the local Lorentz factor. The ratio (\ref{adi}) highlights that the adiabatic approximation becomes reliable for long wavelengths for $a'/ma^2 \ll 1$.

  In the  representation (\ref{solwkb})  it follows that the spinors can be written compactly as
  \be U_s(\vec{k},\eta) = N\,e^{-i\int^{\eta}\omega_k(\eta') d\eta'}\,\left( \begin{array}{c}
                                     (\Omega +M )\, \xi_s\\
                                    k  \, s \, \xi_s
                                  \end{array}\right)\,,  \label{Uad}\ee

 \be V_s(-\vec{k},\eta) = N\,e^{i\int^{\eta}\Omega^{*}_k(\eta') d\eta'}\left( \begin{array}{c}
                                      -k  \, s\,\xi_s\\
                                    (\Omega^* +M )  \,   \xi_s
                                  \end{array}\right)\,,  \label{Vad}\ee with $N$ a normalization constant. The orthogonality conditions $U^\dagger_s U_{s'}= 0, V^\dagger_s V_{s'}=0$ for $s\neq s'$ and  $U^\dagger_s \,V_{s'} =0$ for all $s,s'$ are evident. Furthermore, using the equations (\ref{realOm},\ref{ImagOm}) it is straightforward to   show that
\be \frac{d}{d\eta} \Big( U^\dagger_s\,U_s\Big) =0 ~~;~~ \frac{d}{d\eta} \Big( V^\dagger_s\,V_s\Big) =0 \,, \label{constno}\ee therefore normalizing the spinors $U^\dagger_s U_{s'}= \delta_{s,s'}= V^\dagger_s V_{s'} $  it follows that
\be |N|^2 \, e^{2\int^\eta \Omega_I(\eta')d\eta'}\, \Big[\Omega^2_R + \Omega^2_I + \omega^2 +2M\Omega_R \Big] =1 \,. \label{norad}\ee Up to an overall constant phase this equation yields

 \be N = \frac{e^{-\int^\eta \Omega_I(\eta')d\eta'}}{\Big[\Omega^2_R + \Omega^2_I + \omega^2 +2M\Omega_R \Big]^{1/2}}\,.  \label{noradi}\ee

 Using this result,   the general form of the normalized spinors is given by
\be  U_s(\vec{k},\eta) =  \frac{e^{-i\int^{\eta}\Omega_{R}(\eta') d\eta'}}{\Big[\Omega^2_R + \Omega^2_I + \omega^2 +2M\Omega_R \Big]^{1/2}}\,\left( \begin{array}{c}
                                     (\Omega +M )\, \xi_s\\
                                    k  \, s \, \xi_s
                                  \end{array}\right)\,,  \label{Uadfin}\ee

\be V_s(-\vec{k},\eta) =  \frac{e^{i\int^{\eta}\Omega_{R}(\eta') d\eta'}}{\Big[\Omega^2_R + \Omega^2_I + \omega^2 +2M\Omega_R \Big]^{1/2}}\left( \begin{array}{c}
                                      -k  \, s\,\xi_s\\
                                    (\Omega^* +M )  \,   \xi_s
                                  \end{array}\right)\,.  \label{Vadfin}\ee To leading (zeroth) adiabatic order with $\Omega_R = \omega, \Omega_I =0$ we find
                                  the spinors given in eqns. (\ref{Uspinrdasy},\ref{Vspinrdasy}).

It is straightforward to find the general result (for each polarization $s$) ($\Sigma = \alpha\cdot{\vec{k}}+\gamma^0 M$)
\be V^\dagger(-\vec{k},\eta) \,\Sigma(\vec{k},\eta)\,V(-\vec{k},\eta) = -    U^\dagger(\vec{k},\eta) \,\Sigma(\vec{k},\eta)\,U(\vec{k},\eta) = -\frac{M(\Omega^2_R+\Omega^2_I+\omega^2)+2\omega^2 \Omega_R}{\big[\Omega^2_R+\Omega^2_I + \omega^2 + 2M\Omega_R \big]}\,. \label{contrho}\ee

 \be V^\dagger(-\vec{k},\eta) \,\vec{\alpha}\cdot \vec{k} \,V(-\vec{k},\eta) = -  U^\dagger(\vec{k},\eta) \,\vec{\alpha}\cdot \vec{k}\,U(\vec{k},\eta) = - \frac{2k^2\,\big(\Omega_R+M \big)}{\big[\Omega^2_R+\Omega^2_I + \omega^2 + 2M\Omega_R \big]}\,.  \label{contP}\ee

 These results imply that the adiabatic expansion for both energy density and pressure are  {even} in adiabatic derivatives confirming some results in ref.\cite{rio}.

 Up to second order in the adiabatic expansion we find
\be  V^\dagger(-\vec{k},\eta) \,\Sigma(\vec{k},\eta)\,V(-\vec{k},\eta) = -    U^\dagger(\vec{k},\eta) \,\Sigma(\vec{k},\eta)\,U(\vec{k},\eta) = - \omega\Bigg[1 -\frac{1}{8}\,\Big( \frac{a'}{ma^2}\Big)^2  \, \Big(\frac{k}{\gamma^2\,\omega} \Big)^2 \Bigg] \,, \label{rho2nd}\ee
 \bea V^\dagger(-\vec{k},\eta) \,\vec{\alpha}\cdot \vec{k} \,V(-\vec{k},\eta) &  = &  -  U^\dagger(\vec{k},\eta) \,\vec{\alpha}\cdot \vec{k}\,U(\vec{k},\eta) = -\frac{k^2}{\omega_k}\Bigg\{1- \frac{1}{8\,\gamma^4}\,\Big( \frac{a'}{ma^2}\Big)^2 \, \nonumber \\ & \times & (1+\frac{1}{\gamma})\Big[1-\frac{1}{\gamma}+2(1+ \frac{1}{\gamma})\, \Big( \frac{\gamma-2}{(1+\gamma)^2}\Big)  \Big]
 \Bigg\} \label{pres2nd}
 \eea

where $\gamma\equiv \sqrt{1+(k/ma)^2}$ is the local Lorentz factor. These results   agree  with those in ref.\cite{rio}.

For $a \simeq a_{eq} \simeq 10^{-4}$ it follows that
\be \Big( \frac{a'}{ma^2}\Big)^2 \simeq \frac{10^{-54}}{(m/\mathrm{eV})^2}\,. \label{secord}\ee

At leading (zeroth) adiabatic order $\Omega_R = \omega;\Omega_I =0$ the spinors are instantaneous eigenstates of the instantaneous conformal time Dirac Hamiltonian $\Sigma(\vec{k},\eta)$, namely,
\be  \Sigma(\vec{k},\eta)\,U(\vec{k},\eta) =  \omega \,U(\vec{k},\eta)~~;~~  \Sigma(\vec{k},\eta)\,V(\vec{k},\eta) =  -\omega \,V(\vec{k},\eta)\,. \label{zeroeigen}\ee Consequently the interference terms $\rho_{int}$ (\ref{rhoint}) vanish identically at zeroth adiabatic order, however, this is not the case for the interference terms in the pressure.

\end{document}